\documentclass[usenatbib,usegraphicx]{mn2e}
\usepackage{amssymb}
\usepackage{verbatim}
\usepackage{xcolor}
\newcommand{\halfspace}{\hspace{1pt}}

\newcommand{\MSun}{\mathop{\rm M_\odot}\nolimits}

\newcommand{\kms}{\mathop{\rm km \ s^{-1}\,}\nolimits}
\newcommand{\Lya}{Ly$\alpha$}

\newcommand\HI{{\hbox{H\halfspace$\rm \scriptstyle I$}}}

\newcommand\HeII{{\hbox{He\halfspace$\rm \scriptstyle II$}}}

\newcommand\lsim{~\lower.5ex\hbox{$\buildrel < \over \sim$}~}
\newcommand\gsim{~\lower.5ex\hbox{$\buildrel > \over \sim$}~}
\usepackage{color}

\title[Gas around galaxy haloes: methodology comparisons]{Gas around
  galaxy haloes: methodology comparisons using hydrodynamical
  simulations of the intergalactic medium}
       
\author[Avery Meiksin, James S. Bolton, Eric R. Tittley]{
        Avery Meiksin$^{1}$\thanks{E-mail:\ A.Meiksin@ed.ac.uk (AM)},
        James S. Bolton$^{2}$, Eric R. Tittley$^{1}$\\
        $^{1}$SUPA\thanks{Scottish Universities Physics Alliance},
	Institute for Astronomy, University of Edinburgh,
        Blackford Hill, Edinburgh\ EH9\ 3HJ, UK\\
        $^{2}$School of Physics and Astronomy, University of Nottingham,
        University Park, Nottingham\ NG7\ 2RD, UK}

\shortcites{}

\begin{document}

\date{Accepted . Received ; in original form }
\pagerange{\pageref{firstpage}--\pageref{lastpage}} \pubyear{2014}
\maketitle
\label{firstpage}

\begin{abstract}

  We perform cosmological simulations of the intergalactic medium
  (IGM) at redshift $z\sim 3$ using the numerical
  gravity-hydrodynamics codes \texttt{GADGET-3} and \texttt{Enzo} for
  the purpose of modelling the gaseous environments of galaxies. We
  identify haloes in the simulations using three different
  algorithms. Different rank orderings of the haloes by mass result,
  introducing a limiting factor in identifying haloes with observed
  galaxies. We also compare the physical properties of the gas between
  the two codes, focussing primarily on the gas outside the virial
  radius, motivated by recent \HI\ absorption measurements of the gas
  around $z\sim2$--$3$ galaxies. The internal dispersion velocities of
  the gas in the haloes have converged for a box size of 30 comoving
  Mpc, but the centre-of-mass peculiar velocities of the haloes have
  not up to a box size of 60 comoving Mpc. The density and temperature
  of the gas within the instantaneous turn-around radii of the haloes
  are adequately captured for box sizes 30~Mpc on a side, but the
  results are highly sensitive to the treatment of unresolved, rapidly
  cooling gas, with the gas mass fraction within the virial radius
  severely depleted by star formation in the \texttt{GADGET-3}
  simulations. Convergence of the gas peculiar velocity field on large
  scales requires a box size of at least 60~Mpc. Outside the
  turn-around radius, the physical state of the gas agrees to 30
  percent or better both with box size and between simulation
  methods. We conclude that generic IGM simulations make accurate
  predictions for the intergalactic gas properties beyond the halo
  turn-around radii, but the gas properties on smaller scales are
  highly dependent on star formation and feedback implementations.

\end{abstract}


\section{Introduction}
\label{sec:Intro}

The gaseous environments of forming galaxies are expected to be a
maelstrom of activity. Gaseous flows into dark matter haloes feed
galaxies with material for creating stars. The resulting supernovae
drive outflows that may impede or disrupt the inflow. These outflows
may have several consequences on the growth of galaxies and their
gaseous environments. They may regulate star formation in the
galaxies, open up pathways for the release of ionizing photons that
contribute to the metagalactic photoionization background, and
possibly distribute metals over intergalactic scales.

Observational evidence for outflows in moderate redshift ($1.5\lsim
z\lsim3$) galaxies has been mounting for over a decade. Spectral
measurements of star-forming galaxies reveal blue-shifted metal
absorption lines, sometimes accompanied by enhanced blue Balmer
emission wings or red-shifted \Lya\ emission
\citep[e.g.][]{1996ApJ...462L..17S, 1997ApJ...486L..75F,
  2000ApJ...528...96P, 2001ApJ...554..981P, 2009ApJ...692..187W,
  2010ApJ...717..289S, 2010MNRAS.402.1467Q,
  2011ApJ...733..101G}. While mass flow rate estimates are fraught
with uncertainties, the absorption and velocity signatures suggest
outflow rates comparable to the star formation rates of the galaxies,
with a large reservoir of cool gas built up by the outflows in the
circumgalactic region \citep{2010ApJ...717..289S,
  2011ApJ...733..101G}.

On the other hand, evidence for cold, inflowing gas has been less
forthcoming. Inflows may either arise from cosmological accretion onto
the haloes, or by returning gas carried outward earlier by
winds. Detection of the inflow patterns around moderate redshift
($z=2-3$) galaxies has, however, recently been provided by velocity
measurements of the \Lya\ optical depth of neutral hydrogen in the
vicinity of galaxies \citep{2012ApJ...751...94R,2012ApJ...750...67R},
extending from circumgalactic scales out to several comoving Mpc.

Feedback in the form of winds driven by supernovae have long been
suspected of regulating the inflow and outflow of gas around galaxy
haloes \citep{1971ApJ...170..241M, 1974MNRAS.169..229L}. The
gravitational influence of dark matter will favour mass loss via winds
from low mass haloes over large, and may be responsible for the
distinction between dwarf and normal galaxies
\citep{1986ApJ...303...39D}. The ram pressure of cosmological
accretion onto sufficiently massive haloes may even trap a wind within
the turn-around radius, where the gas has decoupled from the Hubble
expansion and is inflowing, leading to renewed infall and star
formation \citep{2004ApJ...613..159F}.

Given the variety of complex, non-linear physical mechanisms at play
in the gaseous environments of galaxy haloes, cosmological
hydrodynamical simulations are widely used to model the observational
data. However, although many simulations with winds have been
performed, the basic driving mechanism of the winds, whether by
pressure or by momentum, and basic parameters like the mass loading
factor are still unknown \citep[e.g.][]{1989ApJ...337..141M,
  2003MNRAS.339..289S, 2005ApJ...618..569M, 2008MNRAS.387..577O,
  2012MNRAS.426..140D, 2013MNRAS.429.1922C}. A further complication is
that some winds may be driven by Active Galactic Nuclei (AGN). As a
consequence, even when models predict a wind will be present, it is
unclear how far the wind will travel, how much mass it carries, and
even whether it will escape the galaxy into the intergalactic medium
(IGM) or fall back onto the galaxy.

In contrast, numerical simulations have been very successful at
predicting the properties of the lower density, largely quiescent
intergalactic gas probed by the \Lya~forest to high accuracy
\citep{2009RvMP...81.1405M}. Despite lacking sub-grid implementations
for winds, these models can nevertheless prove useful as a tool for
interpreting and calibrating the absorption signatures of the gas
around galaxies. To do so, however, it is necessary to establish how
accurately the observable properties of the IGM may be predicted in
these models, and to determine at which scales commonly used IGM
simulation methodologies are no longer sufficient to model complex
gaseous halo environments.

The purpose of this paper is to describe in detail the capacity and
limitations of simulations specifically designed for modelling the IGM
when applied to the extended gaseous environment of galaxies. This
work will focus on moderate redshift galaxies in the range $2<z<3$ in
particular, for which the surrounding gas has been probed by
\HI\ absorption line studies along lines of sight to background
quasars, as in the Very Large Telescope Lyman-break galaxy redshift
survey \citep{2011MNRAS.414...28C} and the Keck Baryonic Structure
Survey \citep{2012ApJ...750...67R}. The latter authors in particular
divide the gaseous environment of galaxies into three zones:\ a
circumgalactic zone within 300~kpc (proper) of the galaxy, which
approaches the turn-around radius of the galaxy haloes; an
intermediate zone between 300~kpc and 2~Mpc (proper), and the ambient
IGM at larger distances. The most massive haloes may also be useful
for modelling the environments of quasar hosts, which show evidence
for large amounts of cool gas \citep{2006ApJ...651...61H,
  2013ApJ...762L..19P}. In this work we shall demonstrate that IGM
simulations are able to converge on the physical properties of the gas
outside the circumgalactic zone, specifically beyond the turn-around
radii of the gas accretion onto the haloes, but require a detailed
star formation prescription to model accurately the gas within. Any
disagreement between the simulation predictions of \HI\ properties and
those measured beyond the turn-around radius would suggest winds
influence gas outside the circumgalactic zone. We are examining this
topic in a companion paper.

In order to demonstrate this, the two key factors we investigate in
the IGM simulations are the uncertainty in the simulated halo masses
associated with the observed galaxies, and the numerical agreement of
the physical properties of the gas as computed by differing simulation
methodologies. We use two widely used gravity-hydrodynamics codes for
this purpose:\ \texttt{GADGET-3}, an updated version of the publicly
available code \texttt{GADGET-2} \citep[last described
  by][]{2005MNRAS.364.1105S}, and \texttt{Enzo}
\citep{2014ApJS..211...19B}. In the first half of this paper we
investigate the selection of dark matter haloes. No single
halo-finding algorithm of the many in the literature is overall better
than the rest; at some level the identification of haloes, and in
particular the masses assigned to them, depend on arbitrary choices of
technique. The issues involved have received wide attention in the
literature for low redshift haloes \citep[e.g.][]{2002ApJS..143..241W,
  2009ApJ...692..217L, 2011ApJ...732..122B, 2012MNRAS.423.1200O,
  2013MNRAS.435.1618K, 2013MNRAS.433.1230W, 2013arXiv1310.3740K,
  2014arXiv1402.4461V}, but less so for the redshifts of interest
here, at $2<z<3$ \citep{2007MNRAS.374....2R, 2008ApJ...688..709T,
  2013MNRAS.433.1230W}. We adopt three different methods and assess
the differences in the properties of the haloes identified. In the
second half of the paper, we compare the properties of the gas
surrounding the haloes as computed by \texttt{GADGET-3} and
\texttt{Enzo}.

All results are presented for a flat $\Lambda$CDM universe with the
cosmological parameters $\Omega_m=0.28$, $\Omega_bh^2=0.0225$ and
$h=H_0/100~\kms=0.70$, representing the total mass density, baryon
density and Hubble constant, respectively. The intial matter power
spectrum in the simulations has a spectral index $n=0.96$, and is
normalized to $\sigma_{8h^{-1}}=0.82$, consistent with the 9-year {\it
  Wilkinson Microwave Anisotropy Probe} ({\it WMAP}) data
\citep{2013ApJS..208...19H}.  

This paper is organised as follows. In the next section we describe
the cosmological simulations used in this work. The halo catalogues
constructed from these results are discussed in
section~\ref{sec:haloes}, and the properties of the gaseous
environments of the haloes are presented in
section~\ref{sec:haloprops}. Readers interested primarily in the
comparison of the properties of gas around galaxy haloes in the
simulations may skip directly to section~\ref{sec:haloprops}.  Our
conclusions are summarised in section~\ref{sec:conclusions}. A short
appendix contains technical details on the convergence requirements
and appropriate parameter choices when identifying dark matter haloes
in the simulations.

\section{Numerical simulations}
\label{sec:sims}

\begin{table*}
  \centering
    \begin{minipage}{180mm}
      \begin{center}
    \caption{Summary of the simulations performed in this work.
        The columns, from left to right, list the simulation name, the
        box size in comoving Mpc, the number of resolution elements in
        the simulation, the code used for the run, the star formation
        prescription and whether or not the model includes supernovae
        driven winds. }
    \begin{tabular}{l c c c c r}
      \hline\hline
      Name       & Box size     & Resolution & Method    & Star
      formation & Winds       \\
                 & [Mpc] & elements \\
      \hline
      G30qLy$\alpha$  &      30  &  2$\times$512$^3$ &  {\texttt{GADGET-3}}  &
      qLy$\alpha$ & N \\
      G30sfnw         &      30  &  2$\times$512$^3$ &  {\texttt{GADGET-3}}  &
      SH03 & N \\
      G30sfw          &      30  &  2$\times$512$^3$ &  {\texttt{GADGET-3}}  &
      SH03 & Y \\
      \hline
      E30\_512  &           30  &  512$^3$ &  {\texttt{Enzo-2}}  &
      none & N \\
      E60\_1024 &           60  & 1024$^3$ &  {\texttt{Enzo-2}}  &
      none & N \\
      \hline
    \end{tabular}
    \label{tab:sims}
\end{center}
  \end{minipage}
\end{table*}

\subsection{Cosmological hydrodynamics codes}

We use two widely used gravity-hydrodynamics codes in this analysis,
one particle based and the other grid based. The particle based code
\texttt{GADGET-3}, which is an updated version of the publicly
available code \texttt{GADGET-2} \citep{2005MNRAS.364.1105S}, uses
Smoothed Particle Hydrodynamics (SPH) to solve the fluid equations and
a particle-based tree algorithm for gravity. By contrast,
\texttt{Enzo} \citep{2014ApJS..211...19B} solves the fluid equations,
including the gravity of the baryons, on a mesh, and the dark matter
gravitational forces on the top level grid using a hybrid
particle-mesh (PM) scheme. An extension of the method is to adapt the
mesh resolution as necessary using adaptive mesh refinement. Tests
show the success in resolving dark matter haloes below the top grid is
sensitive to the means of triggering the refinements
\citep{2005ApJS..160....1O, 2008CS&D....1a5003H}. We consider only
unigrid (top level grid) simulations here, in keeping with the typical
approach used for IGM analyses.

The numerical simulations were performed in boxes of size 30~Mpc
(comoving) on a side using {\texttt{GADGET-3}} and {\texttt{Enzo}},
v.2.1.1. As we focus on moderate redshift haloes, the runs were
performed down to $z=2$ only. The {\texttt{GADGET-3}} simulations were
run with $512^3$ gas particles and $512^3$ cold dark matter
particles. The {\texttt{Enzo}} simulation was run with a top-level
only grid of $512^3$ mesh zones and $512^3$ cold dark matter
particles. The dark matter particle mass in these simulations is
$m_c=6.4\times10^6\, \MSun$, and the gas particle mass (or mean gas
mass per grid zone) is $m_g=1.3\times10^6\,\MSun$. These simulation
parameters ensure good convergence on the statistics of the \Lya\
forest at $z\sim 2-3$ \citep{2004MNRAS.350.1107M,
  2005MNRAS.357.1178B}. As a test of convergence on the properties of
the gas surrounding the haloes with box size, we also perform a second
\texttt{Enzo} simulation in a 60~Mpc box with $1024^3$ mesh zones and
$1024^3$ cold dark matter particles. We note the standard initial
conditions generation routines differ between \texttt{GADGET-3} and
\texttt{Enzo}. We stress we have not sought to generate identical
initial conditions for the two codes, but rather to examine
differences between the overall code methodologies. In this sense we
are not performing head-to-head code comparisons, but rather seeking
the regime of agreement between two different generic IGM simulations
using two widely used codes, examining in particular the gaseous
environments of galactic mass haloes as computed by the simulations.

Both the {\texttt{Enzo}} and {\texttt{GADGET-3}} computations used
identical background photoionization histories and atomic rates for
the heating and cooling, as discussed in \citet{2007MNRAS.380.1369T},
except for adopting the \HI\ electron excitation and collisional
cooling rate of \citet{1991ApJ...380..302S}. We use the
photoionization and photoheating rates of \citet{2012ApJ...746..125H},
which include contributions from both galaxies and quasars. The \HeII\
heating rate was modified to reproduce the IGM temperature evolution
of \citet{2011MNRAS.410.1096B} for $\gamma=1.3$. The UV background is
switched on at $z=15$ and is applied in the optically thin limit. The
codes were also modified to solve the non-equilibrium ionization rate
equations.

Any computation of the IGM also requires a means of avoiding the high
computational expense incurred by following rapidly cooling gas. Our
focus in this study is on gas outside the galaxy haloes, within which
the bulk of this cooling occurs. Nevertheless, the treatment of
rapidly cooling gas will impact on the baryonic material throughout
the vicinity of a halo. We therefore investigate the effects of
different means of treating rapidly cooling gas on the simulation
results in some detail. This enables us to establish the region around
galaxies over which the means of treating unresolved rapidly cooling
gas no longer affects predictions for the intergalactic gas, allowing
reliable predictions to be made for comparison with observations.

The \texttt{Enzo} simulations bypass the problem of rapidly cooling
gas by simply not spatially resolving regions that would produce rapid
gas cooling. This is possible because the Jeans length of the IGM well
exceeds the scales of rapid gas cooling in collapsed haloes. Because
of its Lagrangian nature, however, \texttt{GADGET-3} will inevitably
track regions of high gas density and rapid cooling, so that some
means of gas removal is necessary. We implement gas removal using two
methods. The first is a simplified prescription, \lq quick Ly$\alpha$'
(labelled G30qLy$\alpha$ below), which converts all gas particles with
an overdensity $\Delta>1000$ and gas temperature $T<10^5$~K into
collisionless particles (categorized as `star' particles in the code),
significantly speeding up the computation
\citep{2004MNRAS.354..684V}. We emphasize that this prescription is a
computational trick and is not meant to represent actual star
formation. A second method (G30sfnw) implements the multi-phase star
formation prescription of \citet{2003MNRAS.339..289S}. Although
designed to include winds, we turn off the wind option to compare with
the \texttt{Enzo} results. Lastly, we also perform a simulation
(G30sfw) using the \texttt{GADGET-3} supernovae driven wind model of
\citet{2003MNRAS.339..289S}, as an exploration of the impact a wind
may have on the properties of the gas surrounding the haloes compared
with the non-wind case. This model assumes a wind velocity of
$484\rm\,km\,s^{-1}$, where each galaxy has a mass outflow rate twice
its star formation rate, and the energy of the wind is equal to the
energy released by supernovae. The simulations are summarised in
Table~\ref{tab:sims}.

\subsection{Halo finding}

Central to any statistical predictions of the properties of galaxies
is the selection of simulated haloes meant to represent them. Various
statistics are available to match haloes in a simulation volume to
observed galaxies. The most straightforward is abundance matching.
This involves simply matching simulated haloes to observed galaxies
according to the rank order of the simulated halo masses and an
observed extensive property of a galaxy, such as total luminosity or
velocity dispersion. Another method compares the clustering strengths
of galaxies and simulated dark matter haloes. No method is perfect,
however, as the definition of a halo must contain some element of
arbitrariness. Allowing for feedback in the form of radiation and
galactic winds further complicates any matching procedure. Which
definition relates best to observed galaxies is a matter of contention
which likely will not be resolved without a more complete theory of
galaxy formation. Many aspects of these issues have been explored in
the literature \citep[e.g.][]{1988ApJ...327..507F,
  2004ApJ...609...35K, 2006MNRAS.371.1173V, 2010ApJ...710..903M,
  2013ApJ...770...57B, 2014arXiv1404.3724S}.

In this study we focus our discussion on dark matter haloes in the
mass range $11<\log_{10}(M/\MSun)<12$, although we shall consider
trends outside this range as well. Based on clustering strength and
luminosity-limited number counts, \citet{2012ApJ...752...39T} estimate
the galaxies in the sample of \citet{2012ApJ...751...94R} from the
Keck Baryonic Structure Survey occupy haloes with a minimum total mass
(dark matter and baryons) of $\log_{10}(M/\MSun) > 11.7\pm0.1$ and a
median total mass of $\log_{10}(M/\MSun) = 11.9\pm0.1$. These halo
masses are also consistent with those inferred from the amount of
\HI\ absorption arising from the circumgalactic gas of the galaxies
\citep{2013MNRAS.433.3103R}.

We use two different particle based methods to select the haloes in
the simulations:\ Friends-of-Friends (FoF) \citep{1982ApJ...259..449P,
  1984MNRAS.206..529E, 1985ApJ...292..371D}\footnote{We use a
  publicly-available code at
  http://www-hpcc.astro.washington.edu/tools/fof.html .} and HOP
\citep{1998ApJ...498..137E}. We also introduce a new method that
selects haloes based on the density field interpolated onto a grid.
The FoF algorithm joins all particles within a given fixed distance of
one another, usually set according to the mean distance between
particles. A disadvantage of FoF is that it sometimes links together
sets of particles that to the eye would be regarded as separate haloes
joined by a bridge. The HOP algorithm is designed to overcome this
difficulty, forming groups by jumping to particles in ever denser
neighbourhoods until no denser neighbour may be found. The groups tend
to be more isolated than found using the FoF algorithm, although an
allowance is made to join separate clumps if bridged by regions above
a given density threshold. The HOP algorithm shares with FoF the
advantage of being scale-free, but relies on more parameters. In
practice, however, it is the outer density threshold for inclusion in
a group that is the primary parameter that defines the group catalog.
For FoF, we adopt the standard linking length of 0.2 the mean
inter-particle separation. For HOP, we take $\delta_{\rm outer}=80$,
which we find gives good agreement with the FoF halo numbers. The
remaining parameters are set in accordance with the recommendations in
the documentation accompanying the HOP source code.\footnote{We take
  $N_{\rm dens}=64$, $N_{\rm hop}=16$, $N_{\rm merge}=4$ with
  $\delta_{\rm outer}=\delta_{\rm saddle}/2.5 = \delta_{\rm peak}/3$,
  using the quantities defined in \citet{1998ApJ...498..137E}.}

The force softening scale for the {\texttt{GADGET-3}} runs is 1.4~kpc
(comoving). Since {\texttt{Enzo}} uses the particle-mesh method, the
force resolution is limited to two grid zones, or 118~kpc (comoving),
adequate for resolving the Jeans length of the photoionized gas. The
minimum virialized\footnote{Haloes with central dark matter densities
  exceeding the virialization density will be referred to as \lq
  virialized'; this is not meant to imply the haloes are necessarily
  in virial equilibrium. The virial mass $M_V$ is the mass contained
  within the virial radius, defined here as the radius within which
  the average dark matter overdensity is $18\pi^2$ relative to the
  background dark matter density.} halo mass achievable in the
{\texttt{Enzo}} computation in a single cell is thus
$18\pi^2(m_c+m_g)=1.4\times10^{9}\,\MSun$. Allowing for a minimum of
27 cells to resolve a virialized halo in the gridded density field
corresponds to a minimum mass of $3.7\times10^{10}\,\MSun$. Much lower
mass haloes are achievable in the {\texttt{GADGET-3}} run (and the
{\texttt{Enzo}} run, if using adaptive mesh refinement) in principle,
but not necessarily if the haloes are to avoid being under-resolved or
over-relaxed in IGM simulations. Further details on this point may be
found in the appendix.

\section{Halo catalogues}
\label{sec:haloes}
\subsection{Baryon mass fraction in haloes}
\label{subsec:baryonfraction}

\begin{figure}
\scalebox{0.47}{\includegraphics{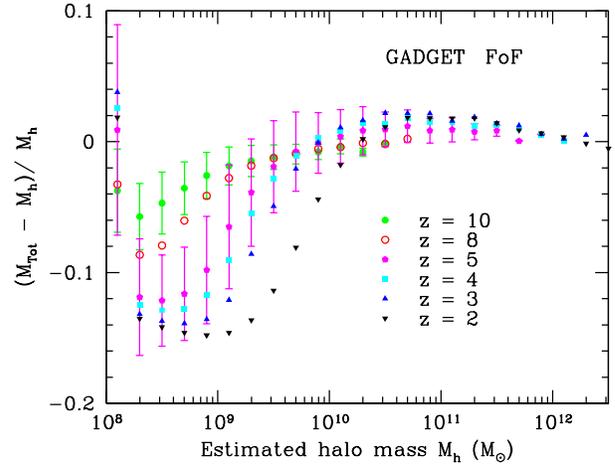}}
\vspace{-1.0cm}
\caption{Difference between total halo mass $M_{\rm Tot}$ and rescaled
  dark matter halo mass $M_h$ assuming the cosmic mean mass ratio of
  baryons to dark matter. Shown at $z=2$, 3, 4, 5, 8 and 10 for haloes
  found using Friends-of-Friends in a {\texttt{GADGET-3}} run. The error
  bars indicate the $1\sigma$ spread in differences at $z=5$ and 10.}
\label{fig:FoFDMass}
\end{figure}

\begin{figure}
\scalebox{0.65}{\includegraphics{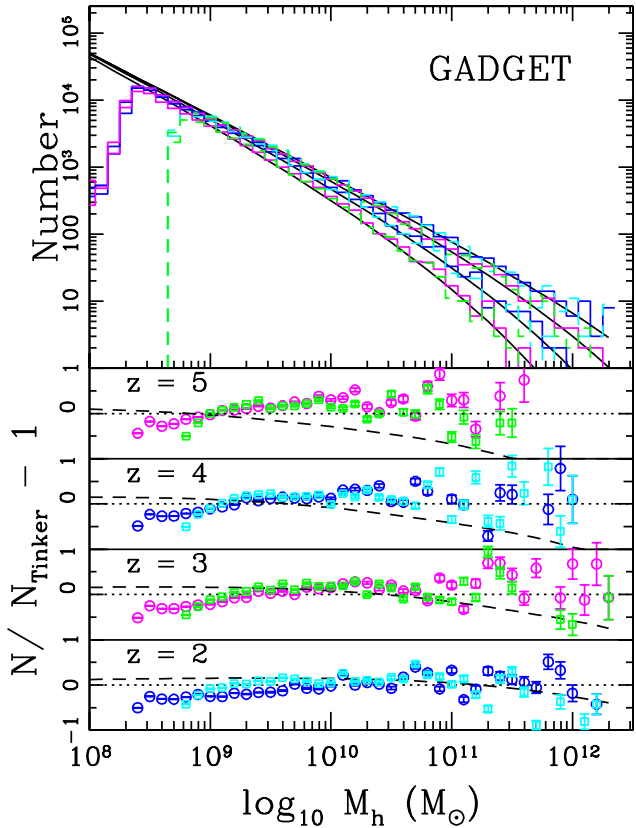}}
\vspace{-1.5cm}
\caption{Halo mass distribution function at $z=2$, 3, 4 and 5, for the
  G30qLy$\alpha$ \texttt{GADGET-3} simulation. {\it Upper panel:}
  Number of haloes found using FoF (solid; magenta and blue online,
  alternated for clarity) and HOP (dashed; green and cyan online,
  alternated for clarity), along with the expected number using the
  fit of \citet{2008ApJ...688..709T} (black), with the sets of curves
  increasing at the high mass end from $z=5$ to $z=2$. The simulation
  results correspond to the total halo mass scaled from the dark
  matter component, assuming the cosmic mean mass ratio of baryons to
  dark matter. {\it Lower panels:} The fractional deviation of
  simulation halo counts from the model of
  \citet{2008ApJ...688..709T}, for the FoF haloes (circles) and HOP
  haloes (squares). The error bars are Poisson. The dashed lines show
  the expected counts using the fitting formula of
  \citet{2007MNRAS.374....2R}.
}
\label{fig:gadgetHMF}
\end{figure}

\begin{figure}
\scalebox{0.65}{\includegraphics{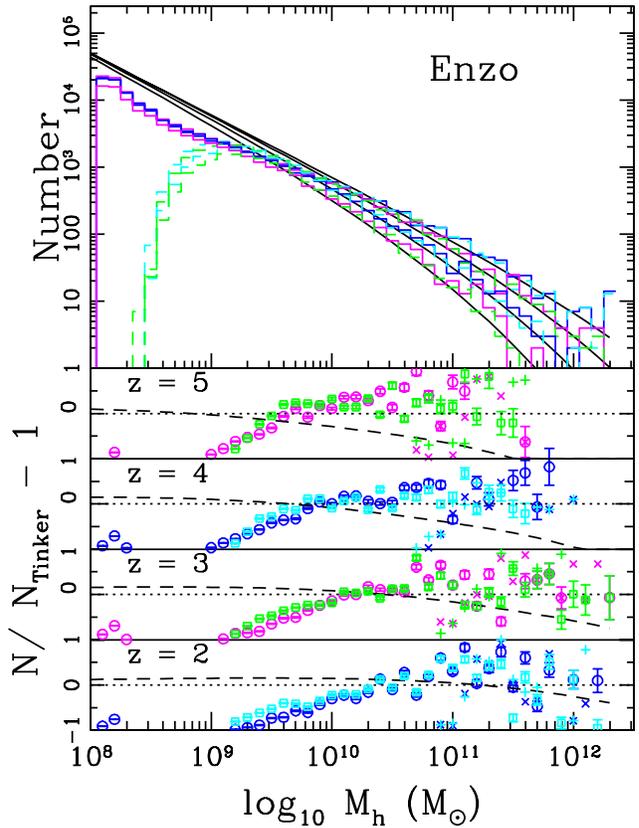}}
\vspace{-1.5cm}
\caption{As for Fig.~\ref{fig:gadgetHMF}, except now showing the halo
  mass distribution function for the E30\_512 \texttt{Enzo} run. Note
  the estimated minimal resolvable halo mass in this unigrid
  simulation is $4\times10^{10}\,\MSun$. In the lower panels the halo
  masses for \lq profiled' virialized haloes are shown by `+'s for FoF
  haloes, and by `x's for HOP haloes (see text for further details).}
\label{fig:enzoHMF}
\end{figure}

We now turn to describing the properties of the dark matter haloes in
our simulations, before going on to discuss the gaseous environments
of the haloes in section~\ref{sec:haloprops}.

As most of the literature on haloes uses dark matter only simulations,
the total halo mass (dark matter and baryons) is often scaled from the
dark matter component assuming a uniform mass ratio of baryons to dark
matter equal to the mean cosmic value, $\Omega_{\rm b}/(\Omega_{\rm
  m}-\Omega_{\rm b})$. We first test this assumption in
Fig.~\ref{fig:FoFDMass} by applying FoF to the \texttt{GADGET-3}
simulation G30qLy$\alpha$ for a range of redshifts. (Note the results
are nearly identical for the G30sfnw simulation, which we do not show
here.) The rescaled halo masses $M_h$ are scaled from the dark matter
assuming the cosmic baryon to dark matter mass ratio, and the actual
halo masses $M_{\rm Tot}$, given by the combined mass of the dark
matter, gas and star particles in the model, are found to agree
closely over most of the halo mass range.

At the low mass end, however, discrepancies arise, with the actual
mass systematically smaller than the rescaled mass, although with wide
scatter. This difference arises primarily from heating by the UV
background, increasing the thermal gas pressure and so impeding the
inflow of the gas \citep[e.g.][]{1992MNRAS.256P..43E,
  2008MNRAS.390..920O}. The difference is small at $z=10$ and 8, but
by $z=6$ the discrepancy exceeds 10 percent, with the range in
discrepant masses systematically increasing with decreasing
redshift. By $z=2$, the discrepancy exceeds 10 percent for haloes less
massive than $4\times10^9\,\MSun$, corresponding to a characteristic
temperature of $T\simeq50\times10^3$~K, comparable to the temperature
of reionized intergalactic gas, including the enhanced heating rate as
the UV metagalactic ionization background hardens, adiabatically
compressed to virial densities. The effect of this redshift dependent
baryonic physics suggests the dimensionless mass function shape is not
universal to a precision better than 15 percent at the low halo mass
end at these redshifts, even in the absence of supernovae
feedback. With this mass discrepancy at the low mass end in mind,
unless stated otherwise, halo masses in the remainder of this paper
refer to values rescaled from the dark matter component assuming a
uniform baryon to dark matter ratio at the cosmic mean value. As we
focus primarily on halo masses in the mass range
$11<\log_{10}(M/\MSun)<12$, this should be a reasonable approximation.
  
\subsection{Halo mass function}
\label{subsec:hmf}

The halo mass functions obtained by running FoF and HOP on the
G30qLy$\alpha$ {\texttt{GADGET-3}} dark matter particles are shown in
Fig~\ref{fig:gadgetHMF}. We compare the numbers of haloes found with
the fitting formula of \citet{2008ApJ...688..709T} for overdensity
$\Delta=200$ haloes\footnote{The halo mass is defined as the mass
  contained within a spherical surface centred on the halo and having
  an average internal overdensity 200 times the cosmic mean
  density. The halo masses using this definition well match those
  using FoF with $b=0.2$ \citep{1996MNRAS.281..716C,
    2008ApJ...688..709T}.} (solid black curves in upper panel),
allowing for redshift dependent coefficients. Since this fitting
formula was based on spherical overdensity haloes and the redshift
dependence was limited to $0<z<2.5$, we also compare with the fitting
formula of \citet{2007MNRAS.374....2R} in the lower panels of the
figure. This is based on haloes with masses
$10^5-10^{12}\,h^{-1}\,\MSun$, identified over the redshift interval
$0<z<30$ using FoF with a linking length of $b=0.2$. The expected
counts were generated using the {\texttt{genmf}} fitting formula code
provided by \citet{2007MNRAS.374....2R}, adjusted to our cosmological
parameters. Since we search for haloes using only the dark matter
component, as noted earlier the total mass of the haloes is found by
allowing for a baryon component at the cosmic mean ratio of baryons to
dark matter. This matches the halo mass definitions of
\citet{2008ApJ...688..709T} and \citet{2007MNRAS.374....2R}.

The FoF halo mass distribution in Fig~\ref{fig:gadgetHMF} agrees very
closely with the fitting formula of \citet{2008ApJ...688..709T},
within the scatter, for halo masses $M>2\times10^9\,\MSun$. The
scatter sometimes exceeds the Poisson errors, based on the number of
haloes found in a mass bin, but excess scatter is expected from
large-scale structure, especially for the rarer haloes.  At $z=4$ and
5, the halo numbers continue to agree well with the
\citet{2008ApJ...688..709T} fitting formula, but deviate from the Reed
fitting formula, which differs from the number of haloes we obtain by
as much as $\sim50$ percent at the high mass end, suggesting the
fitting formula coefficients may not extend well to the different
cosmological and power spectrum parameters we used, which more closely
agree with those of simulations included in the
\citet{2008ApJ...688..709T} analysis.

The halo mass function for the HOP haloes is remarkably similar to the
FoF halo mass function, within the scatter. There is a $\sim20$
percent excess for $10^9<M<10^{10}\,\MSun$ at $z=2$, bringing the
numbers more closely in line with the \citet{2008ApJ...688..709T} halo
mass function. None the less, the differences in the counts suggests
the algorithms are not always identifying the halo masses
consistently. We return to this point below.

The halo mass functions from the E30\_512 {\texttt{Enzo}} simulation,
shown in Fig~\ref{fig:enzoHMF}, also generally agree with the
\citet{2008ApJ...688..709T} mass function for haloes with
$M>2\times10^{10}\,\MSun$ for $z=2$ and 3, although with considerable
scatter. This mass threshold is comparable to the minimum mass for
achieving a virialization density in 15 contiguous cells. The
agreement extends down to $4\times10^9\,\MSun$ at $z=5$. The mass
functions found from the FoF and HOP algorithms generally agree, but
vary at the $\sim20$ percent level for halo masses below
$10^{10}\,\MSun$. The more conservative halo resolution requirement of
27 contiguous mesh zones requires a minimum virialized halo mass of
$M>3.7\times10^{10}\,\MSun$, and we take this to be representative of
the resolvable halo mass in the simulation.

Finally, we also \lq profile' the haloes in the E30\_512
{\texttt{Enzo}} simulation by first constructing spherical density
profiles centred on the densest dark matter point in a halo, and then
computing the virial mass of the halo by scaling from the dark matter
mass to account for the baryonic component. The profiled results for
virialized haloes are shown as `x's for the FoF haloes, and `+'s for
the HOP haloes in the lower panels of Fig.~\ref{fig:enzoHMF}. The
number of virialized haloes falls off abruptly below $10^{11}\,\MSun$
relative to the \citet{2008ApJ...688..709T} mass function. As for the
{\texttt{GADGET-3}} haloes, the numbers between the FoF and HOP haloes
do not precisely match.

\begin{figure}
\scalebox{0.65}{\includegraphics{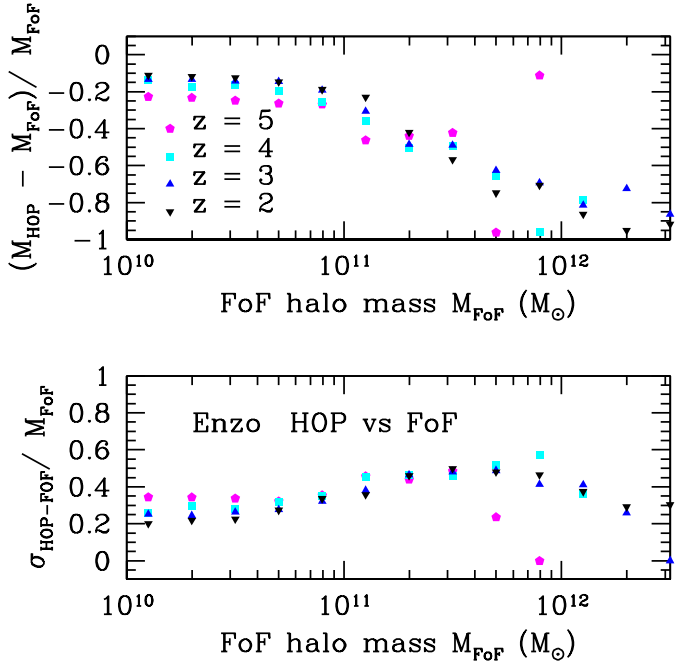}}
\vspace{-1.2cm}
\caption{The difference between the FoF and HOP halo masses from the
  E30\_512 {\texttt{Enzo}} run as a function of FoF halo mass. The
  halo match is based on finding the HOP halo which lies within the
  virial radius of a given FoF halo with the most similar mass. The
  comparison is shown at $z=2$, 3, 4 and 5. {\it Upper panel:} Mean
  difference in the halo masses. {\it Lower panel:} The standard
  deviation in the difference between halo masses.}
\label{fig:enzoFoFHOP}
\end{figure}

\subsection{Minimum halo mass consistently identified by FoF and HOP}
\label{subsec:minhalomass}

The principal source of the discrepancy between the halo mass
distributions produced by different halo finding algorithms for well
resolved haloes is generally not that different haloes are identified
(although this may occur in unusually complex regions of massive
mergers). Instead, different algorithms will typically assign
different masses to the same haloes
\citep[e.g.][]{2008MNRAS.385.2025C}. However, provided these
algorithms agree on the halo centres about which any subsequent radial
density profiles are constructed, rank ordering the haloes by their
estimated virial mass should provide a stable basis for comparing with
observed galaxies; this approach will be largely independent of the
means used for identifying the haloes. In this sub-section, we
therefore examine the minimum halo mass for which FoF and HOP, as
applied to the \texttt{Enzo} data, produce identical virial
masses. Note that since \texttt{GADGET-3} resolves halo density
profiles to smaller scales than \texttt{Enzo}, the corresponding
minimum halo mass of \texttt{GADGET-3} haloes will be smaller. We
therefore confine our discussion of the minimum consistently
identified halo mass to the \texttt{Enzo} data only.

We first compare the masses of individual haloes identified by both
FoF and HOP by one-to-one matching HOP and FoF haloes identified in
the E30\_512 \texttt{Enzo} simulation.  We achieve this by searching
for the nearest HOP halo within the virial radius of a FoF halo. A
comparison between the halo masses is shown in
Fig.~\ref{fig:enzoFoFHOP}. The HOP halo mass generally agrees well (to
within around 20 per cent) with the FoF halo mass below
$10^{11}\,\MSun$. At higher masses, however, the HOP halo masses are
increasingly low compared with the corresponding FoF mass, consistent
with HOP's breaking up chains of particles that FoF links together
into the same halo. This shows that, while the halo finders identify
the same peaks, they associate somewhat different particles to the
resulting haloes. As a consequence, the FoF and HOP haloes do not
maintain the same rank ordering by mass, with a spread in mass
difference of around 20--60 percent, as shown in the lower panel of
Fig.~\ref{fig:enzoFoFHOP}. In the absence of a more precise definition
of halo mass, this partially undermines the use of rank ordering when
associating simulated haloes with observed galaxy properties.

The virial mass (which we obtain by profiling the haloes in the manner
described previously) offers a much better definition for this
purpose, since the rank ordering is preserved by the different halo
finders above a minimum halo mass. To demonstrate the stability of the
virialized halo masses against the choice of halo finder for
sufficiently massive haloes, we one-to-one match virialized haloes
found by FoF and HOP. At $z=3$, nearly one-third of the virialized FoF
haloes have no matching virialized HOP halo located within the virial
radius of the FoF halo. Conversely, nearly one-quarter of virialized
HOP haloes have no matching virialized FoF halo. Almost all the
unmatched haloes have masses below $2\times10^{11}\,\MSun$. For the
remaining majority of virialized haloes, the total halo masses within
the virial radius computed from the density profiles centred on the
density peak found by either halo finder agree almost exactly. Thus
the virial masses of essentially all virialized haloes with masses
above $2\times10^{11}\,\MSun$ agree, whether identified using FoF or
HOP. Should future surveys extend measurements of circumgalactic gas
to smaller halo masses, a grid code like \texttt{Enzo} would then
require higher spatial resolution than we have used, either using a
finer top-grid or an adaptive mesh, going beyond the standard
requirements for an IGM simulation.

\begin{figure}
\scalebox{0.65}{\includegraphics{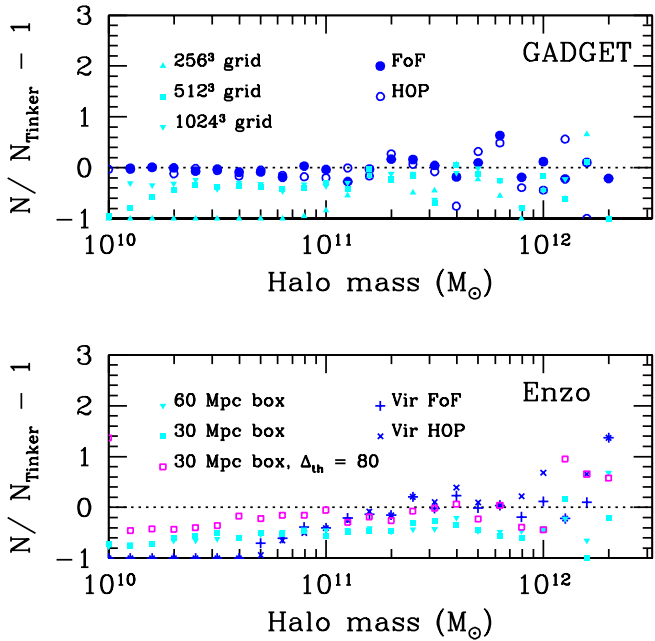}}
\vspace{-1.2cm}
\caption{Differential number counts of haloes at $z=3$ identified
  using a gridded density field halo finder (GHF) based on the dark
  matter density field with a threshold density set at the
  virialization density, relative to the \citet{2008ApJ...688..709T}
  fitting function. {\it Upper panel:} Results for the G30qLy$\alpha$
  \texttt{GADGET-3} density field gridded onto meshes with $256^3$
  (cyan triangles), $512^3$ (cyan squares) and $1024^3$ (cyan inverted
  triangles) cells. For comparison, the counts of haloes found using
  FoF (blue filled circles) and HOP (blue open circles) finders are
  also displayed. {\it Lower panel:} Results for the haloes found from
  the dark matter density field for two \texttt{Enzo} simulations of
  fixed resolution, but with comoving box sizes of 30~Mpc (cyan
  squares) and 60~Mpc (cyan inverted triangles). Also shown are the
  results for haloes found above a threshold overdensity of 80
  (cf. $18\pi^{2}$) in the 30~Mpc box (magenta open squares). For
  comparison, the counts of virialized haloes found using FoF (blue
  `+'s) and HOP (blue `x's) are also shown.
}
\label{fig:griddedHMF}
\end{figure}

\subsection{Grid-based halo finder}
\label{subsec:GHF}

An alternative approach to particle based halo finders is to identify
haloes on the dark matter density grid. For large simulations, this
has the advantage of requiring far less data to be saved, particularly
for a hydrodynamical grid code. Moreover, since the gravitational
force is computed on a grid in mesh codes, haloes found from the
gridded density field will more faithfully reflect the resulting mass
concentrations.

Motivated by these considerations, we have developed a gridded density
field halo finder (GHF) similar to the search for spherical
overdensities in $N$-body data \citep{1992ApJ...399..405W,
  1994MNRAS.271..676L}, although our method is based on the local
density rather than a mean internal density, and so tracks the
filamentary structure of overdense regions. Specifically, the method
grows haloes by building them up in concentric shells about density
peaks, with the following procedure:\ 1.\ Identify all the density
peaks of the gridded density field, and rank order them from highest
to lowest. 2.\ Working down the list from the highest peak, search
among the next nearest layer of mesh cells for those with densities
above a given threshold overdensity $\Delta_{\rm th}$. 3.\ If a cell
incorporated on the list appears on the list of density peak cells,
remove it from the list of peaks. (It is assumed incorporated into the
halo with a higher density peak.) 4.\ If the fraction of cells more
overdense than $\Delta_{\rm th}$ exceeds a given value $p_{\rm th}$,
repeat step 2 extending to the next layer; otherwise cease growing the
halo and go to the next density peak on the list. Choosing
$\Delta_{\rm th}=178$ will grow approximately spherical virialized
haloes.

The resulting halo counts are shown in Fig.~\ref{fig:griddedHMF} for
the G30qLy$\alpha$ \texttt{GADGET-3} simulation (upper panel) and
E30\_512 and E60\_1024 \texttt{Enzo} simulations (lower panel). The
\texttt{GADGET-3} dark matter particles are gridded onto meshes with
$256^3$, $512^3$ and $1024^3$ cells. The halo counts lie 30--50
percent lower than predicted by the \citet{2008ApJ...688..709T} halo
mass function, with the deficit increasing towards lower mass
haloes. The agreement improves at the low mass end with increasing
mesh resolution, but otherwise the counts are largely insensitive to
the regridding resolution. Haloes on a $512^3$ ($256^3$) grid are
recovered to 30--50 percent down to $2\times10^{10}\,\MSun$
($2\times10^{11}\,\MSun$).

The number counts of haloes found from the gridded dark matter density
field in the \texttt{Enzo} $512^3$ 30~Mpc volume simulation likewise
lie systematically low compared with \citet{2008ApJ...688..709T}, by
about 30--60 percent. Using the density field from a second
\texttt{Enzo} simulation in a 60~Mpc box with the same grid resolution
as for the 30~Mpc box simulation, and setting $\Delta_{\rm th}=178$,
provides comparable halo numbers to the 30~Mpc box, as shown in
Fig.~\ref{fig:griddedHMF}. The halo numbers are reasonably well
converged, so box size does not account for the differences. The
discrepancy may be attributed in part to the difference in the extents
of the haloes. Lowering the overdensity threshold from $\Delta_{\rm
  th}=178$ to 80 provides much better agreement with
\citet{2008ApJ...688..709T}. As shown in the appendix, lowering this
threshold increases individual halo masses, putting them into better
agreement with the masses of the matching FoF haloes.

\subsection{Observational prediction uncertainties due to uncertainty in halo mass assignments}
\label{subsec:mass_uncertainty}

Finally, we note that the sensitivity of halo mass rankings to the
halo finding algorithm introduces uncertainty into the assignment of
halo masses to observed galaxies if abundance matching is used as a
basis \citep[see e.g. ][for several references to the literature on
  abundance matching]{2014arXiv1404.3724S}. As we have discussed here,
one way to approach the problem is to use only haloes well resolved
within their virial radii, and then rank them by some fixed criterion
like virial mass. Since different halo finders mostly identify the
same structures when well resolved, the masses about the halo centres
will generally agree. For this reason we restrict our analysis in the
next section to haloes with $11<\log_{10}(M/\MSun)<12$. How successful
this approach is at matching observed galaxies, however, remains an
open question.

There are also a large number of lower mass haloes which do not have
well resolved virial cores in our simulations that may still be useful
for statistical analyses. The uncertainty in the masses of these
haloes will give rise to an uncertainty in any predicted properties of
observed galaxies and their environments. An approximate means of
estimating the impact of the uncertainty in halo mass on the
dispersion in a predicted property is to average it over a Gaussian
distribution, allowing that any given halo finding algorithm may err
in the assignment of halo mass $M_h$ with a standard deviation
$\sigma(M_h)=\beta M_h$. We have found typical values of
$\beta=0.2-0.4$ in our analysis above (e.g.,
Fig.~\ref{fig:enzoFoFHOP}). For a property that may be approximated as
a power law in mass, $f(M_h) \sim M_h^\alpha$, it is then
straightforward to show that in the limit $\vert
\alpha(\alpha-1)\vert\beta\ll1$, the mean is only quadratically
biased, $\langle f\rangle/ \langle f\rangle_{\beta=0} - 1 \simeq
(1/2)\alpha(\alpha-1)\beta^2$, while the relative standard deviation
is $\sigma_f/\langle f\rangle \simeq \vert\alpha\beta\vert$.  As an
example, the estimated velocity dispersion of a halo, $v_{\rm rms}\sim
(GM_h/r_V)^{1/2}\sim M_h^{1/3}$, will be biased low by 1 percent, with
a relative spread of 10 percent, for $\beta=0.3$. Since the actual
halo mass probability distribution may have a broad tail, this
approach may conservatively be regarded as providing a lower limit on
the uncertainty.

\section{Intergalactic medium properties around haloes}
\label{sec:haloprops}

\subsection{Halo peculiar velocities}
\label{sec:halovpec}

\begin{figure}
\scalebox{0.65}{\includegraphics{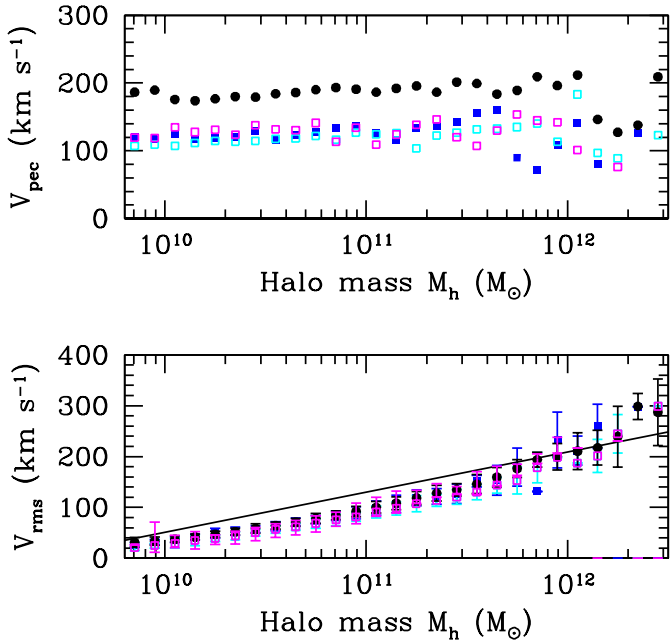}}
\vspace{-1.2cm}
\caption{Convergence of the halo peculiar velocities (top panel) and
  internal velocity dispersion of the gas (lower panel) for the
  \texttt{Enzo} simulations E30$\_$512 (filled squares; blue) and
  E60$\_$1024 (filled circles; black) at $z=3$, as a function of halo
  mass. Also shown are the values for halo masses using a lower
  overdensity threshold of 80 in E30$\_$512 (open squares; cyan). The
  results for haloes in the \texttt{GADGET-3} simulation G30sfnw (open
  squares; magenta) agree well with the results for the corresponding
  \texttt{Enzo} haloes. The solid line in the lower panel shows
  $v_{\rm circ}/2^{1/2}$, where $v_{\rm circ}$ is the circular
  velocity at the virial radius.
}
\label{fig:egvpecvrms}
\end{figure}

In this section, we now turn to analysing the properties of the gas
around galaxy haloes with total masses $11<\log_{10}(M/\MSun)<12$,
corresponding to the haloes of galaxies with measured \HI\ absorption
in their environments. These haloes also tend to be relatively
isolated, permitting clearly delineated radial profiles of the gas
propertires to be constructed.

We first examine the convergence of the gas peculiar velocity
field. Because the peculiar velocity power spectrum peaks on scales in
excess of 100 Mpc (comoving), this is not expected to fully converge
in our 30-60 Mpc boxes on large scales. In Sec.~\ref{sec:radprofs}
below we address the radial scale over which the peculiar velocity
field of a halo converges.

The convergence of the halo peculiar velocity and internal velocity
dispersion $v_{\rm rms}$ of the gas at $z=3$ is shown in
Fig~\ref{fig:egvpecvrms} for GHF haloes with $\Delta_{\rm th}>178$ for
the E30$\_$512 and E60$\_$1024 \texttt{Enzo} and the G30sfnw
\texttt{GADGET-3} simulations. The \texttt{GADGET-3} data are binned onto a
$512^3$ mesh to match the \texttt{Enzo} spatial resolution. Doubling
the box size of the \texttt{Enzo} simulation from 30 to 60~Mpc
(comoving) nearly doubles the peculiar velocities of the haloes,
showing they have not converged. The halo peculiar velocity is
independent of halo mass, showing the haloes behave as test particles
in large scale flows. The internal velocity dispersion of the
\texttt{Enzo} haloes, in contrast, is well converged with box
size. The values for the \texttt{GADGET-3} haloes agree with those for
the \texttt{Enzo} haloes for the corresponding mass bin, as expected
if both codes are producing the same structures for a given halo
mass. Comparison with the halo circular velocity at the virial radius,
$v_{\rm circ}=(GM_h/r_v)^{1/2}$, shows that the gas is dynamically
cool, with $v_{\rm rms}\lsim v_{\rm circ}/2^{1/2}$.

\subsection{Circumgalactic gas properties}
\label{sec:halogas}

\begin{figure}
\scalebox{0.55}{\includegraphics{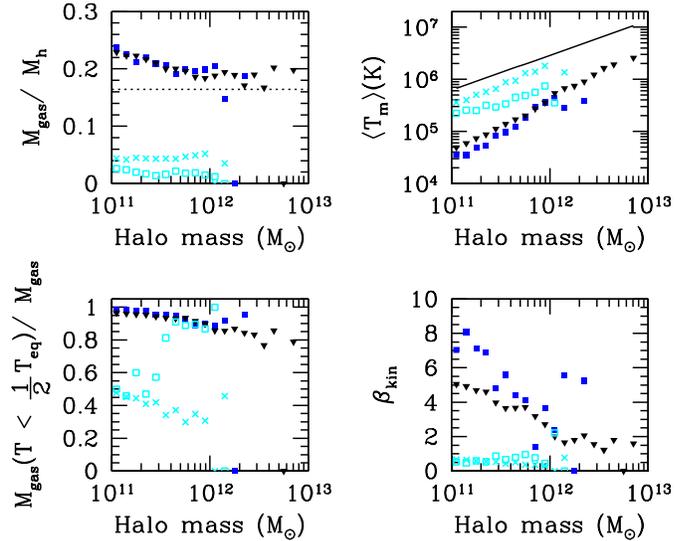}}
\vspace{-1.2cm}
\caption{Thermal and kinetic properties of the gas within the virial
  radius of haloes at $z=3$. Results shown for the \texttt{Enzo}
  simulations E30$\_$512 (blue filled squares) and E60$\_$1024 (black
  filled inverted triangles). Also shown are results for the
  \texttt{GADGET-3} simulation G30sfnw with the gas interpolated onto
  a $512^3$ cell grid (cyan open squares) and the {\texttt{GADGET-3}}
  wind simulation G30sfw (cyan crosses). Clockwise from the top left,
  the panels show the gas mass fraction (the dotted line displays the
  cosmic mean), the mean mass-weighted temperature, the ratio
  $\beta_{\rm kin}$ of the gas kinetic to thermal energies and the gas
  mass fraction with a temperature less than half the halo
  equipartition temperature. The solid line in the upper right panel
  is the predicted post-shock temperature as a function of halo mass.
}
\label{fig:vir_props}
\end{figure}

The thermal and kinetic properties of the gas within the virial radius
of haloes at $z=3$ are shown in Fig.~\ref{fig:vir_props}. The results
are averages over all haloes in mass bins of width
$\Delta\log_{10}M_h=0.1$. The values shown for the \texttt{Enzo} runs
(blue and black symbols) test the convergence of the halo internal gas
properties with box size for 30 and 60~Mpc (comoving) boxes. Results
for the corresponding \texttt{GADGET-3} runs with star formation both
without a wind (cyan squares) and with a wind (cyan crosses) in the
30~Mpc box, gridded onto a $512^3$ mesh to match the \texttt{Enzo}
30~Mpc box, are shown for comparison.

The gas mass fraction for the \texttt{Enzo} haloes is well converged
with box size. For $M_h>10^{12}\,\MSun$, the gas mass fraction lies
just above the cosmic mean value ($\Omega_b/ \Omega_m \simeq 0.164$),
increasing towards lower masses, until 50 percent over-abundant for
$M_h=10^{11}\,\MSun$ haloes. By contrast, star formation in the
\texttt{GADGET-3} simulation G30sfnw leaves behind only a small
fraction of the baryons within the virial radius in the form of gas,
the remainder having been converted into stars. Adding wind feedback
in the G30sfw simulation balances the gas density at somewhat higher
values.

The mean mass-weighted temperature of the gas is defined by
$(3/2)k\langle T_m\rangle/\mu m_{\rm H} = E_{\rm th}/ M_{\rm gas}$,
where $E_{\rm th}$ is the total thermal energy of the gas mass $M_{\rm
  gas}$ within the virial radius, $\mu$ is the mean molecular weight
for a fully ionized hydrogen and helium gas and $m_{\rm H}$ is the
mass of a proton. The temperature is well converged for the
\texttt{Enzo} simulations, as shown in Fig.~\ref{fig:vir_props},
although the convergence worsens for the lower mass haloes. The
temperature of the more rarefied gas in the \texttt{GADGET-3}
simulations is considerably higher. For an adiabatic shock, the
post-shock temperature of a halo of mass $M_h$ collapsing at redshift
$z$ is $T_{\rm shock}\simeq72.1(1+z)(M_h/10^6\,\MSun)^{2/3}$
\citep{2011MNRAS.417.1480M}. The temperature in the \texttt{GADGET-3}
non-wind simulation G30sfnw lies at about one-third this limit,
suggesting radiative losses have been moderately effective in cooling
the post-shock gas. Allowing for a wind in simulation G30sfw produces
somewhat higher temperatures. By contrast, the gas in the
\texttt{Enzo} simulations shows considerable cooling, but does not
lead to runaway cooling on the resolution scale of the grid.

Almost all the gas in the \texttt{Enzo} haloes is colder than half the
halo equipartion temperature $T_{\rm eq}$, defined by $(3/2)\langle
k/\mu m_{\rm H}\rangle T_{\rm eq}=GM_h/r_v$. Nearly the same amounts
are found for the \texttt{GADGET-3} haloes in simulation G30sfnw with
halo masses exceeding $4\times10^{11}\,\MSun$. In lower mass haloes,
cooling is less efficient, with only half the gas cooler than half the
equipartition temperature. In the wind simulation G30sfw, the more
massive haloes have a much smaller proportion of cool gas, with only
one third to one half cooler than half the equipartition temperature.

The gas internal kinetic energy of a halo is defined by
$E_K=(1/2)\int\,dV\,\rho_g({\bf v}_{\rm pec}-{\bf v}_h)^2$, where
$\rho_g$ is the gas density, ${\bf v}_{\rm pec}$ is the gas peculiar
velocity and ${\bf v}_h$ is the centre-of-mass peculiar velocity of
the gas in the halo. The ratio $\beta_{\rm kin}=E_K/E_{\rm th}$
indicates the balance between the kinetic and thermal energies of the
gas. For \texttt{Enzo} haloes with $M_h>10^{12}\,\MSun$, the energies
are nearly in equipartition, with the kinetic energy slightly
larger. The ratio increases to factors of several towards the lower
mass haloes. Comparison between the 30 and 60~Mpc boxes suggests
$\beta_{\rm kin}$ is not yet well converged, with the value decreasing
with increasing box size. For the \texttt{GADGET-3} haloes,
$0.5<\beta_{\rm kin}\lsim1$, suggesting that the gas too rarefied to
rapidly cool and make stars reaches equipartition between the kinetic
and thermal energies.

The large differences between the \texttt{Enzo} and \texttt{GADGET-3}
circumgalactic gas properties demonstrate that the behaviour of the
gas may not be reliably computed outwith a specific star formation
model, even before feedback effects are included. If the star
formation efficiency moreover depends on the internal gas kinematic
properties, then accurate predictions for the properties of
circumgalactic gas pose a severe computational challenge, requiring
both high spatial resolution to follow rapidly cooling gas as well as
a large simulation volume to produce accurate gas flow fields.

\subsection{Radial profiles}
\label{sec:radprofs}

\begin{figure}
\scalebox{0.85}{\includegraphics{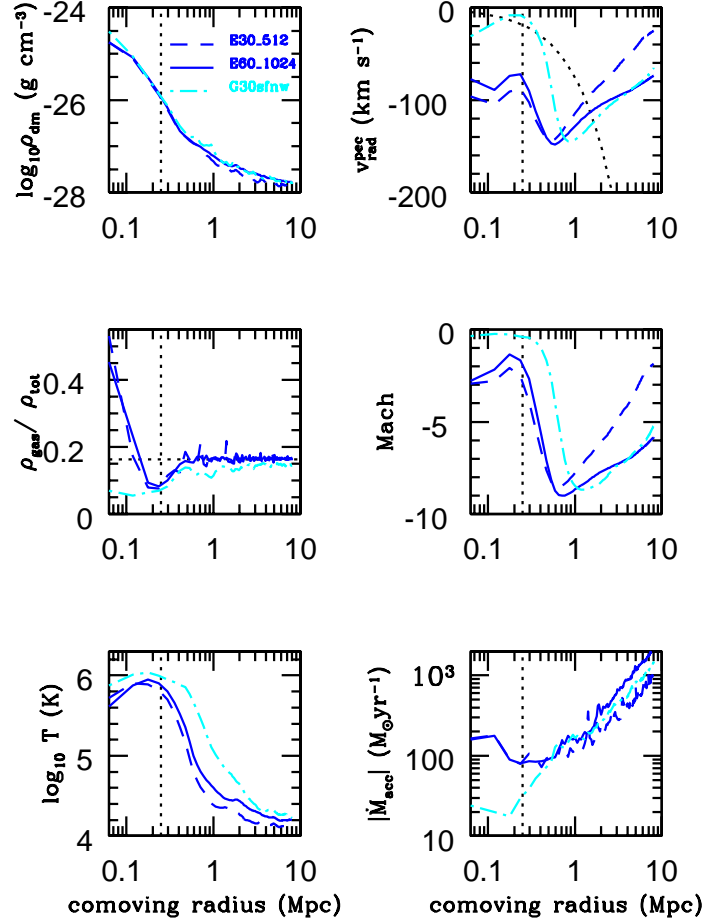}}
\vspace{-2cm}
\caption{Mean radial profiles of halo properties at $z=3$ with total
  masses of $4.5\times10^{11}\,\MSun$, for \texttt{GADGET-3}
  simulation G30sfnw (dot-dashed lines; cyan), and \texttt{Enzo}
  simulations E30$\_$512 (dashed lines; blue) and E60$\_$1024 (solid
  lines; blue). Clockwise from the upper left, the panels display: the
  dark matter density, the gas peculiar velocity, the Mach number, the
  mass accretion rate, gas temperature and the gas density. The
  vertical dotted lines in each panel show the virial radius of the
  haloes. The curved dotted line in the upper right panel shows the
  Hubble expansion (as negative velocity):\ the intersection with the
  peculiar velocity curve indicates the instantaneous turn-around
  radius of the gas in the haloes, located at $r_{t.a.}\simeq 6r_v$.
}
\label{fig:rad_profs}
\end{figure}

\begin{figure*}
\scalebox{0.5}{\includegraphics{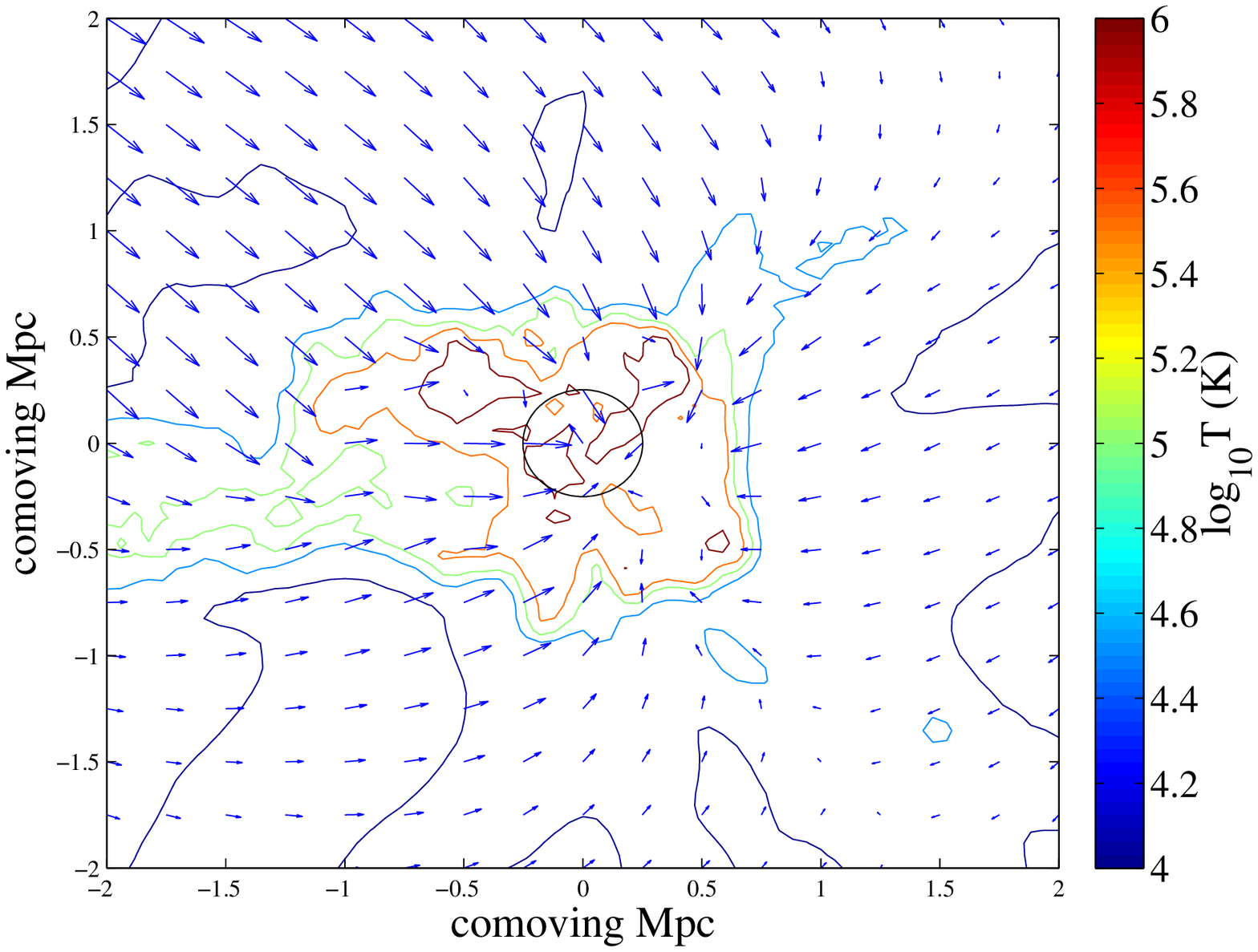}\includegraphics{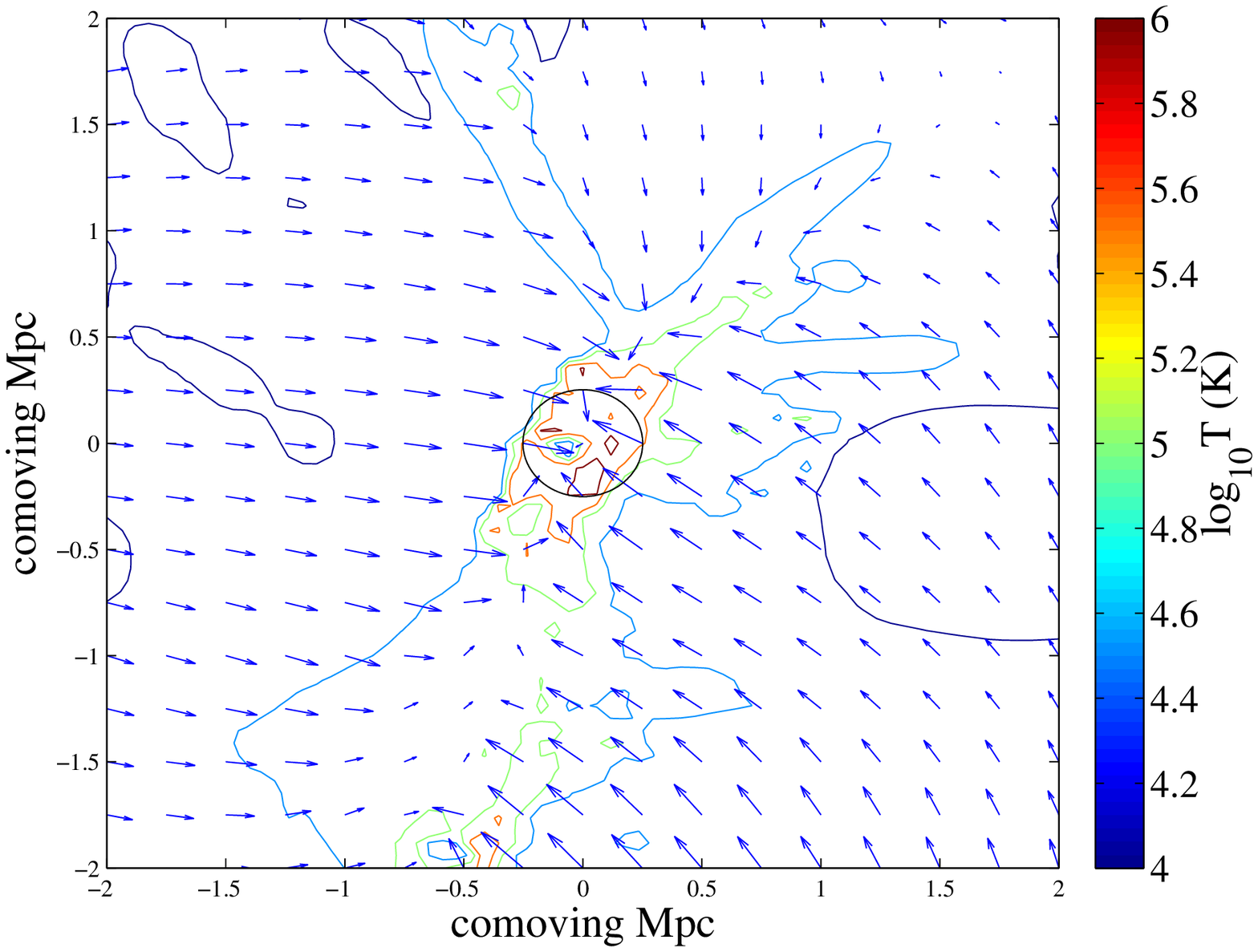}}
\scalebox{0.5}{\includegraphics{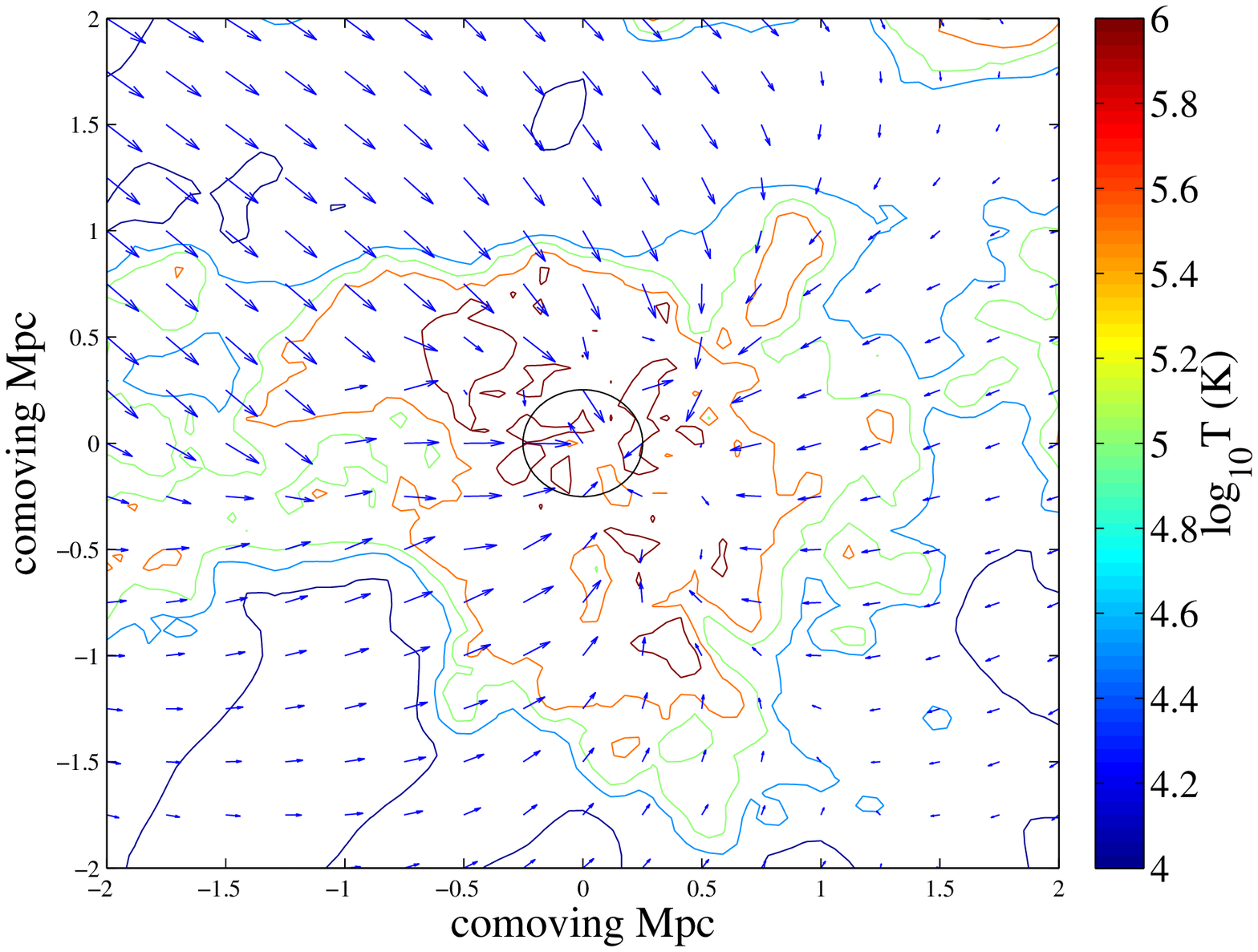}\includegraphics{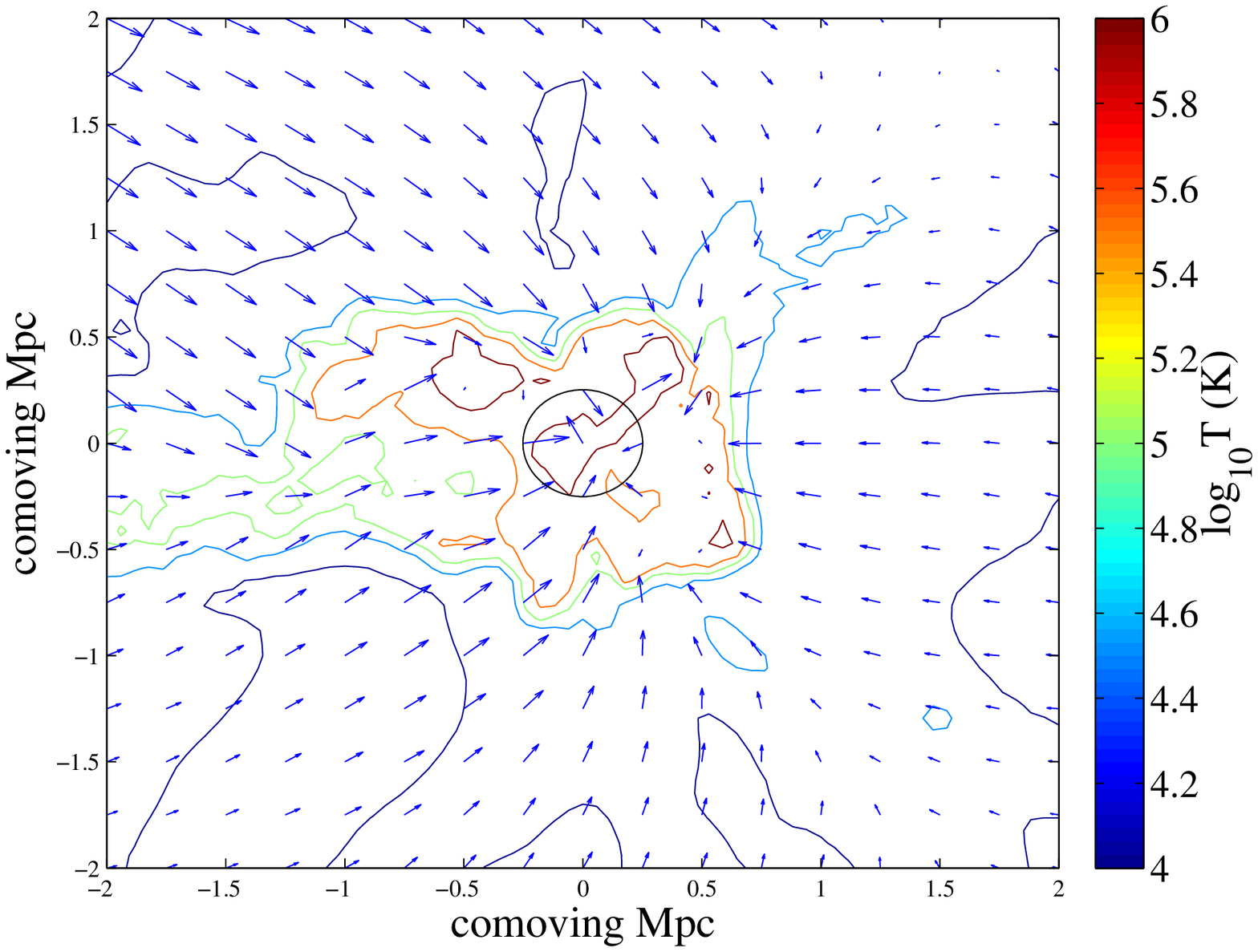}}
\caption{The effects of the prescription for the treatment of
  unresolved, rapidly cooling gas on the gaseous environment of
  haloes, illustrated for representative haloes at $z=3$ with masses
  $\sim5\times10^{11}\,\MSun$ from \texttt{GADGET-3} and
  \texttt{Enzo}. The boxes are 4 comoving Mpc on a side and centred on
  the haloes. Shown are slices of the gas temperature field and
  peculiar velocity flow relative to the halo centres-of-mass. The
  colour bars show $\log_{10} T$. An arrow of one axis tic unit in
  length corresponds to a velocity magnitude of $1000\kms$. The black
  circles indicate the virial radii of the haloes. The panels show
  haloes in simulations, clockwise from the top left, G30sfnw,
  E30$\_$512, G30qLy$\alpha$ and G30sfw. A broad region of hot
  rarefied and kinematically quiescent gas develops around the
  \texttt{GADGET-3} haloes as a result of efficient removal of rapidly
  cooling gas, with enhanced energy input from a wind in simulation
  G30sfw. In the \texttt{Enzo} simulation, gas removal is suppressed
  by the limited spatial resolution, resulting in a more compact and
  kinematically active halo of multiphase gas.
}
\label{fig:haloes}
\end{figure*}

Radial profiles of the dark matter density and gas properties for a
representative halo mass of $4.5\times10^{11}\,\MSun$ are shown in
Fig.~\ref{fig:rad_profs}, for both the \texttt{GADGET-3} and
\texttt{Enzo} haloes in 30~Mpc boxes, as well as \texttt{Enzo} haloes
in the 60~Mpc box. The profiles are averaged over all haloes within a
mass bin of width $\Delta\log_{10}M=0.1$.

A detailed comparison between E30$\_$512 and E60$\_$1024 shows
agreement in the dark matter profiles (upper left panel) within twice
the virial radius to 10 percent, and at 30 percent beyond in the
secondary infall region. The difference may be partly due to low
numbers since there are only 11 haloes in the mass bin. The dark
matter profiles of haloes with somewhat lower masses (not shown) agree
to 15 percent between the two box sizes. The \texttt{GADGET-3} mean
dark matter profile agrees better with the larger box \texttt{Enzo}
simulation. The agreement demonstrates that both \texttt{GADGET-3} and
\texttt{Enzo} are reproducing similar dark matter structures as
identified by halo mass, and that these structures are reasonably well
converged with respect to simulation box size.

\begin{figure}
\scalebox{0.55}{\includegraphics{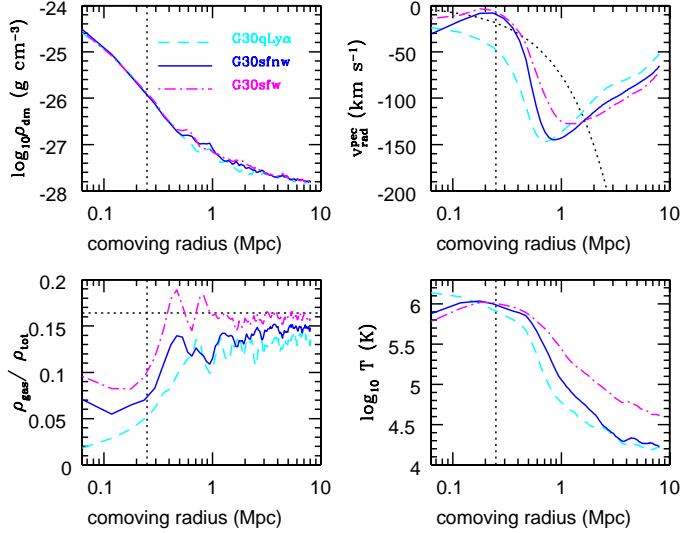}}
\vspace{-1.5cm}
\caption{Mean radial profiles of halo properties for \texttt{GADGET-3}
  haloes at $z=3$ with total masses of $4.5\times10^{11}\,\MSun$, for
  simulations G30sfnw (solid lines; blue), G30qLy$\alpha$ (dashed
  lines; cyan) and G30sfw (dot-dashed lines; magenta). Clockwise from
  the upper left, the panels display: the dark matter density, the gas
  peculiar velocity, gas temperature and the gas density. The vertical
  dotted lines in each panel show the virial radius of the haloes. The
  curved dotted line in the upper right panel shows the Hubble
  expansion (as negative velocity):\ the intersection with the
  peculiar velocity curve indicates the instantaneous turn-around
  radius of the gas in the haloes, located at $r_{t.a.}\simeq 6r_v$.
}
\label{fig:rad_profs_gadget}
\end{figure}

The \texttt{Enzo} simulations have well converged on the physical
state of the intergalactic gas outside the turn-around radii of the
haloes. For radial distances $r>2$~Mpc (comoving), the gas density
profiles agree to within 10 percent and the temperatures to better
than 30 percent (middle and lower left panels). The peculiar inflow
velocity and Mach number agree less well (upper and middle right
panels), although we note somewhat smaller mass haloes show agreement
over $2<r<4$~Mpc to within $\sim30$ percent, but deviate at larger
radii. The mass accretion rate (lower right panel), defined in terms
of the radial peculiar velocity $v^{\rm pec}_{\rm rad}$ as ${\dot
  M}_{\rm acc}=4\pi r^2\rho_{\rm gas}v^{\rm pec}_{\rm rad}$, is
noisier, and shows agreement only at the 50 percent level over
$2<r<4$~Mpc. At larger radii, the mean inflow velocity departs
substantially between the two box sizes, showing poor
convergence. This reflects the non-convergence of the large-scale
peculiar velocity field noted in section~\ref{sec:halovpec}. We note
that the instantaneous turn-around radius of the gas, where the gas
breaks away from the Hubble expansion and begins flowing inward, is
located at $r_{t.a.}\simeq 6r_v$ as shown in the top right panel of
Fig.~\ref{fig:rad_profs}. This is close to the value $\sim4r_v$ for
the self-similar secondary infall of an adiabatic $\gamma=5/3$
collisional gas onto a collapsed dark matter halo in an
Einstein-de~Sitter universe \citep[as inferred from table 8
  of][]{1985ApJS...58...39B}.

While the dark matter profiles agree between the \texttt{GADGET-3} and
\texttt{Enzo} haloes, the gas profiles within the circumgalactic
region show large differences. The \citet{2003MNRAS.339..289S} star
formation prescription in the \texttt{GADGET-3} simulation has removed
most of the baryons within the virial radius from the gas phase. A
wide region of hot gas develops in the \texttt{GADGET-3} haloes
extending over $\sim3$ virial radii, as illustrated in
Fig.~\ref{fig:haloes}. By contrast, in the \texttt{Enzo} haloes the
hot gas region is more compact, with the hot gas component confined to
the inner 1--2 virial radii. A consequence is a lower Mach number for
the accreting gas within the \texttt{GADGET-3} haloes, and a more
quiescent velocity field within and around the haloes. Dense pockets
of cooling gas develop in the \texttt{Enzo} haloes, as illustrated in
Fig.~\ref{fig:haloes}, resembling cold streams
\citep{2003MNRAS.345..349B, 2005MNRAS.363....2K}. An extended warm
stream entering from the left is visible in the \texttt{GADGET-3}
haloes.

Outside the turn-around radius, the \texttt{GADGET-3} and
\texttt{Enzo} results agree well. The baryon fraction in the
\texttt{GADGET-3} haloes is found not to converge to the cosmic mean
value to better than 10 percent by a radial distance of 8~Mpc, as
shown in Fig.~\ref{fig:rad_profs}, suggesting gas removal has been
efficient in the surrounding smaller mass haloes. The
\texttt{GADGET-3} temperature agrees best with the larger box
\texttt{Enzo} simulation, to within 30 percent beyond $r>3$~Mpc. This
may partly be an effect of the gas removal in the \texttt{GADGET-3}
simulation, leaving behind lower density but higher temperature gas.
Achieving better agreement between \texttt{Enzo} and \texttt{GADGET-3}
simulations appears to require a specific model of star
formation:\ the means of dealing with unresolved rapidly cooling gas
has become a limiting factor in the predictive capacity of the
simulations for intergalactic gas near the haloes.

A comparison of the radial profiles of the dark matter density and gas
properties for $4.5\times10^{11}\,\MSun$ haloes from the three
\texttt{GADGET-3} simulations is shown in
Fig.~\ref{fig:rad_profs_gadget}. This directly compares the effect of
different star formation prescriptions on the gas properties. The dark
matter density profiles are essentially unaffected by the mode of gas
removal or the presence of a wind within the virial radius, but
changes of a few tens of percent appear in the secondary infall region
beyond the virial radius. The gas density of the simulations without a
wind lies below the cosmic value out to 8~Mpc, with the quick
Ly$\alpha$ simulation removing gas most efficiently. Invoking a wind
slows the infall velocity of the gas, but has not produced
outflow\footnote{Note also that in the \citet{2003MNRAS.339..289S}
  model, winds remain hydrodynamically decoupled from the gas until
  the gas density is less than 10 per cent the star formation
  threshold density, or if more than 50 Myr has elapsed since the wind
  particle is launched.} in terms of the peculiar velocity. The
position of the outer turn-around radius of the gas remains
unchanged. Less gas is removed from the central regions. A moderate
amount of gas compression occurs beyond the virial radius, with the
gas returning to the cosmic mean baryon density (shown by the
horizontal dotted line) beyond the turn-around radius.

\begin{figure}
\scalebox{0.55}{\includegraphics{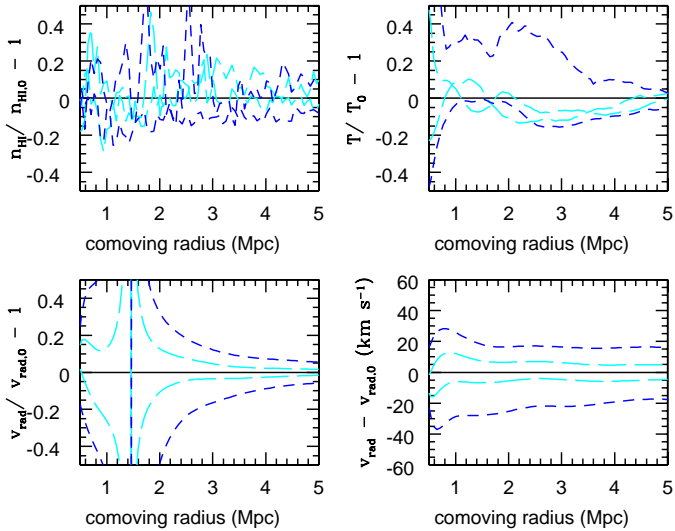}}
\vspace{-1.5cm}
\caption{Percentage differences of halo mean radial profiles from
  $10^{11.55}\,\MSun$ \texttt{Enzo} haloes in 60~Mpc box, for haloes
  with mass offsets $\Delta\log_{10}M_h=\log_{10}M_h-11.55=\pm0.1$
  (cyan long-dashed lines) and $\pm0.3$ (blue short-dashed
  lines). Anti-clockwise from top right, the panels display relative
  differences in the gas temperature, the neutral hydrogen fraction
  and the radial velocity. The absolute velocity difference is shown
  in the lower right panel. The divergence in the relative velocity
  error corresponds to the instantaneous turn-around radius at
  $\gsim1.4$~Mpc.
}
\label{fig:eps_quant_profs}
\end{figure}

Lastly, we note that uncertainty in the simulated halo masses assigned
to galaxies will introduce further uncertainty into the predicted
properties of the intergalactic gas near the haloes. The magnitude of
the uncertainty is illustrated in Fig.~\ref{fig:eps_quant_profs} for a
GHF \texttt{Enzo} halo mass of $10^{11.55}\,\MSun$, showing the
consequences of assigning gas properties corresponding to haloes with
mass offsets of $\Delta\log_{10}M_h=\pm0.1$ and $\pm0.3$. A halo
offset of $\pm0.1$ is representative of the differences in mass
assigned to haloes by the different halo finding algorithms we
used. The resulting neutral hydrogen density $n_{\rm HI}$ and
temperature $T$ differences are 10--20 percent at $r>1.5$~Mpc. A mass
offset of $\pm0.3$ produces differences of 20--40 percent. Comparable
relative differences are found for the total radial velocity except
near the turn-around radius. The absolute radial velocity offsets
range up to $20-30\,\kms$. The relatively modest differences in the
physical properties of the intergalactic gas near haloes (at scales
above the turn-around radius) for a range of halo masses suggest good
tolerance of the predictions to the larger uncertainties in the halo
masses. The converse is that the local circumgalactic gas properties
can provide only a crude estimate of the masses of the central haloes.

\section{Summary and conclusions}
\label{sec:conclusions}

We investigate how well simulations designed to study the IGM
reproduce the physical properties of the gas surrounding galaxy
haloes, motivated by recent observations of the gaseous environments
of redshift $z\sim 2$--$3$ galaxies through \HI\ absorption line
measurements \citep{2010ApJ...717..289S, 2011MNRAS.414...28C,
  2012ApJ...750...67R, 2013ApJ...762L..19P}. To do so, we perform
comparisons of the dark matter and gaseous properties of moderate
redshift haloes, with $2<z<5$, using two different numerical
simulation codes, \texttt{GADGET-3} and \texttt{Enzo}. We have
examined two separate issues, agreement in the halo masses and
abundances, necessary for reliably selecting simulated haloes to
represent observed galaxies, and agreement in the physical properties
of the gas around the haloes. We summarise our results on these topics
separately.

Our main results concerning halo selection are:

1.\ For halo masses exceeding $10^{10}\,\MSun$, rescaling the dark
matter halo mass by the mean cosmic baryon to dark matter density
ratio reproduces the total halo mass (comprised of dark matter and
baryons) to a few percent accuracy. We find, however, that the
rescaling over-estimates the true combined dark matter and baryon mass
of haloes with masses below $10^{10}\,\MSun$ by as much as 15 percent,
with the discrepancy increasing at decreasing redshifts. This
difference is due to the partial loss of gas in the smaller mass
haloes as a result of photoionization heating
\citep[e.g.][]{2008MNRAS.390..920O}.

2.\ Reasonable agreement is obtained between the numbers of FoF and
HOP haloes found in the \texttt{GADGET-3} simulation and the
corresponding \texttt{Enzo} simulation. The halo mass functions agree
with that of \citet{2008ApJ...688..709T} to about 10--30 percent
accuracy over the total halo mass range $10^9<M_h<10^{11}\,\MSun$ for
the \texttt{GADGET-3} haloes and over $10^{10}<M_h<10^{11}\,\MSun$ for
the \texttt{Enzo} haloes, although the halo abundances evolve somewhat
more slowly with redshift for $z>2$ \citep[beyond the redshift range
considered by][]{2008ApJ...688..709T}. There is substantial scatter,
30--50 percent differences from the fitting formula, at higher masses
due to the low numbers of haloes and cosmic variance in our 30 Mpc
(comoving) simulation boxes.

3.\ A one-to-one matching of FoF and HOP haloes in the \texttt{Enzo}
30~Mpc box simulation shows that the HOP halo masses are typically 20
percent smaller than the FoF halo masses for FoF halo mass below
$10^{11}\,\MSun$, nearly independent of redshift. The discrepancy
increases to as much as 80 percent low for $10^{12}\,\MSun$ FoF
haloes. Restricting the analysis to haloes well resolved within their
virial radii, however, shows that the same haloes are identified by
both algorithms and that masses within the virial radii are
identical. To assign simulated halo masses to observed galaxies based
on abundance matching, we thus recommend using only simulated haloes
resolved within their virial radii, preferably by at least 5000
particles to ensure both high resolution and negligible dynamical
over-relaxation, and ranking the haloes by their virial masses.

4.\ Haloes with masses below $2\times10^{11}\,\MSun$ were often not
well resolved within their virial radii by our simulations. The FoF
and HOP halo masses were offset by $\sim20$ percent, and the
dispersion in the mass differences was $\sigma(M_h)=\beta M_h$ with
$\beta\sim0.2-0.4$. FoF and HOP do not preserve the rank ordering of
haloes by mass at this level, undermining the prediction of galaxy
properties to much better than this level of accuracy when halo masses
are assigned to galaxies by abundance matching. In this case, we
suggest a lower limit to the error in a predicted galaxy property may
be estimated by averaging the property and its variance over a
Gaussian distribution in halo mass with $\beta=0.2-0.4$. A halo
property varying as $f(M_h)\sim M_h^\alpha$ will then have a relative
uncertainty of at least $\sigma_f/\langle
f\rangle\simeq\vert\alpha\beta\vert$.

5.\ We introduce a new method for identifying haloes based on a
gridded dark matter density field, similar to the spherical
overdensity method for $N$-body particles. Haloes are found on the
gridded density field by identifying contiguous regions with
overdensity above a given threshold level $\Delta_{\rm th}$. A
practical benefit of the method is that it does not require the
particle data to be saved from a simulation to find haloes, a
particular advantage for massive simulations. For $\Delta_{\rm
  th}=178$, the halo masses are about 30 percent lower than the FoF
masses for the same indentified haloes, but the masses come into good
agreement if the density threshold is lowered to $\Delta_{\rm
  th}=80$. Similarly, for $\Delta_{\rm th}=178$ the \texttt{GADGET-3}
halo counts are offset by $\sim30$ percent below the halo mass
function of \citet{2008ApJ...688..709T}, and by $\sim50$ percent for
the \texttt{Enzo} haloes. Using instead $\Delta_{\rm th}=80$ brings
the counts into good agreement with \citet{2008ApJ...688..709T}.

Our main results concerning the gas properties are:

1.\ \texttt{GADGET-3} and \texttt{Enzo} identify similar halo
structures for a given halo mass for haloes with well resolved virial
cores. For our simulations, these correspond to haloes with masses
exceeding $2\times10^{11}\,\MSun$. The dark matter density profiles
averaged over the haloes agree typically to 10--30 percent over radii
$r_V<r<8$~Mpc (comoving) from the halo centres of mass. The mode of
gas removal, however, affects the dark matter density profile in the
secondary infall region beyond the virial radius by a few tens of
percent. The internal velocity dispersion of the gas in the haloes is
found to agree closely between the \texttt{GADGET-3} and \texttt{Enzo}
haloes. The peculiar velocities of the haloes themselves are poorly
converged with box size, as expected since the velocity power spectrum
has signficant power on scales in excess of 100~Mpc, driving large
scale flows.

2.\ There are pronounced differences in the circumgalactic gas
properties between the \texttt{GADGET-3} and \texttt{Enzo} haloes as a
consequence of the differences in the treatement of unresolved rapidly
cooling gas. The \texttt{GADGET-3} simulation converts most of the gas
into collisionless particles inside the haloes. The mass-weighted
temperature of the remaining gas within the virial radius is
substantially higher than that of the gas in the corresponding
\texttt{Enzo} haloes. A broad high temperature region extending over
2--3 virial radii develops around the \texttt{GADGET-3} haloes. The
\texttt{Enzo} simulations suppress the rapid cooling of gas because of
their restricted spatial resolution. The hot regions of the haloes are
more compact than in the \texttt{GADGET-3} haloes, and develop a
multiphase medium including cooling gas within the virial cores. We
conclude that any predictions of the physical properties of the
circumgalactic gas may be made only within the context of a specific
gas removal prescription.

3.\ Outside the turn-around radii, the gas density and temperature
agree to 30 percent between comoving box sizes of 30 and 60~Mpc, and
to 40 percent between the \texttt{GADGET-3} and \texttt{Enzo}
simulations, without reaching better than 10 percent agreement until
as far out as several turn-around radii. The physical properties of
the gas at these distances may be reliably computed, although the
treatment of rapidly cooling gas is still a limiting factor in the
accuracy of the predictions out to several turn-around radii.

4.\ The wind model we implemented in a \texttt{GADGET-3} simulation
affects the circumgalactic gas, slowing the accretion but not
producing outflow in terms of the peculiar velocity. It increases the
extent of the hot haloes, doubling the gas temperature compared with
the windless model beyond the turn-around radius, while the gas
density converges to the cosmic mean value.

5.\ A halo mass offset of $\pm0.1$ dex compared with a population of
observed galaxies will introduce errors in the predicted neutral
hydrogen density, gas temperature and gas velocities of 10--20 percent
outside the turn-around radii of the haloes. An offset of $\pm0.3$ dex
increases the errors to 20--40 percent. The high tolerance of the
properties of the intergalactic gas near the haloes to the uncertain
halo mass should permit predictions of the \HI\ absorption line
properties of the gaseous environment of galaxies to good
accuracy. Large discrepancies with observations would suggest the
influence of a wind.

We conclude that galactic mass haloes with essentially the same dark
matter properties are reproduced at $2<z<5$ by \texttt{GADGET-3} and
\texttt{Enzo} IGM simulations in 30~Mpc comoving volumes and a
spatially resolved Jeans length. Although the masses assigned to
haloes are sensitive to the halo finding algorithm, the different halo
finding algorithms we consider identify largely the same systems for
halo masses exceeding $\sim4\times10^{10}\,\MSun$, and essentially
identical systems for masses exceeding $\sim2\times10^{11}\,\MSun$.
The physical properties of the intergalactic gas surrounding the
\texttt{GADGET-3} and \texttt{Enzo} haloes with masses exceeding
$\sim2\times10^{11}\,\MSun$ are found to agree to 30--40 percent
beyond the turn-around radii. At smaller radii, the \texttt{GADGET-3}
and \texttt{Enzo} haloes show substantial differences in the gas
density and temperature as a consequence of the differences in the
treatment of rapidly cooling gas on spatially unresolved scales. We
thus conclude that generic IGM simulations are able to make accurate
predictions for the intergalactic gas properties of observed moderate
redshift galaxies beyond the halo turn-around radii, but the
properties of circumgalactic gas are highly dependent on the choices
of star formation and feedback implementation.

\section*{Acknowledgments}
This work used the DiRAC Data Analytic system at the University of
Cambridge, operated by the University of Cambridge High Performance
Computing Service on behalf of the STFC DiRAC HPC Facility
(www.dirac.ac.uk). This equipment was funded by BIS National
E-infrastructure capital grant (ST/K001590/1), STFC capital grants
ST/H008861/1 and ST/H00887X/1, and STFC DiRAC Operations grant
ST/K00333X/1. DiRAC is part of the National
E-Infrastructure. Additional computations were performed on facilities
funded by an STFC Rolling-Grant and consolidated grant. AM thanks
B. Smith for information regarding details of the operations of yt
(http://yt-project.org), used for analysing some of the
{\texttt{ENZO}} simulations, and J. Cohn and M. White for helpful
conversations. JSB acknowledges the support of a Royal Society
University Research Fellowship. ERT is supported by an STFC
consolidated grant. We thank V. Springel for making \texttt{GADGET-3}
available. Computations described in this work were performed using
the {\texttt{Enzo}} code developed by the Laboratory for Computational
Astrophysics at the University of California in San Diego
(http://lca.ucsd.edu).

\bibliographystyle{mn2e-eprint}
\bibliography{apj-jour,ms}

\begin{appendix}
\section{Minimum halo mass in IGM simulations}
\label{ap:minhalomass}

In this appendix, we show the simulations used in this work are
adequate for resolving and selecting galaxy haloes of mass exceeding
$10^{11}\,\MSun$, matching observed galaxies with associated
\HI\ absorption measurements.

We first show the haloes will not be overly dynamically relaxed. A
halo of mass $M_h$ comprised of $N$ particles will have a median
dynamical relaxation time of $t_{\rm rh}\simeq
[0.138N/\log(0.4N)](r_h^3/G M_h)^{1/2}$ \citep{1987degc.book.....S},
where $r_h$ is the half mass radius. In terms of the Hubble time
$t_{\rm H}=2/3H(z)$ for a virialized halo of overdensity $18\pi^2$
compared with the cosmic mean density, the criterion for negligible
two-body relaxation is
\begin{equation}
  \frac{t_{\rm rh}}{t_{\rm
      H}}\simeq\frac{2^{1/2}}{4\pi}\frac{0.138N}{\log(0.4N)}\left[1+\frac{\Omega_v}{\Omega_m(1+z)^3}\right]^{1/2}\gg1,
\label{eq:relax}
\end{equation}
or $N\gg 310$ for $z\gg1$ (and $N\gg136$ at $z=0$). In terms of the
simulations presented here, this corresponds to the halo mass limit
$M_h=N(m_c+m_g)\gg2.4\times10^9\,\MSun$ for $z\gg1$ (and
$M_h\gg1.0\times10^9\,\MSun$ at $z=0$). A safer lower limit to ensure
negligible over-relaxation is $t_{\rm rh}/t_{\rm H}>10$, corresponding
to minimum particle numbers per halo of $N_{\rm min}>4880$ for $z\gg1$
(and $N_{\rm min}>2330$ at $z=0$).  This corresponds to a minimal halo
mass of $M_{h, {\rm min}}=N_{\rm
  min}(m_c+m_g)=3.8\times10^{10}\,\MSun$ to ensure negligible
over-relaxation effects for $z\gg1$. This is comparable to the minimum
resolvable halo mass in the \texttt{Enzo} simulations.

These estimates are similar to those based on convergence tests on
halo properties. Using \texttt{GADGET} simulations of increasing mass
resolution with identical initial conditions,
\citet{2010ApJ...711.1198T} find the masses of FoF haloes identifiable
in different resolution simulations have typical uncertainties of
$\Delta M_h/M_h\sim 1.5/N^{1/3}$, a dependence they attribute to
errors in the number of particles in the halo peripheries. On this
criterion, achieving a halo mass precision of 10 percent requires
$N>3000$ particles, or $M_h>2.3\times10^{10}\,\MSun.$
\citet{2011ApJ...732..122B} suggest halo masses are biased high in
$N$-body simulations and are more accurate if corrected by the factor
$M_c/M=[1.0-0.04(\epsilon/ 650\,{\rm kpc})](1-N^{-0.65})$, where
$\epsilon$ is the force resolution in (comoving) kpc. Applied to our
\texttt{GADGET-3} and \texttt{Enzo} runs with $N=N_{\rm min}$, this
corresponds to a correction by 0.5 and 1 percent, respectively. Since
these are smaller than the accuracy we require, we do not include this
correction. We conclude that the haloes we focus on in this paper
should be free of resolution, force error and over-relaxation
systematics.

\begin{figure}
\scalebox{0.47}{\includegraphics{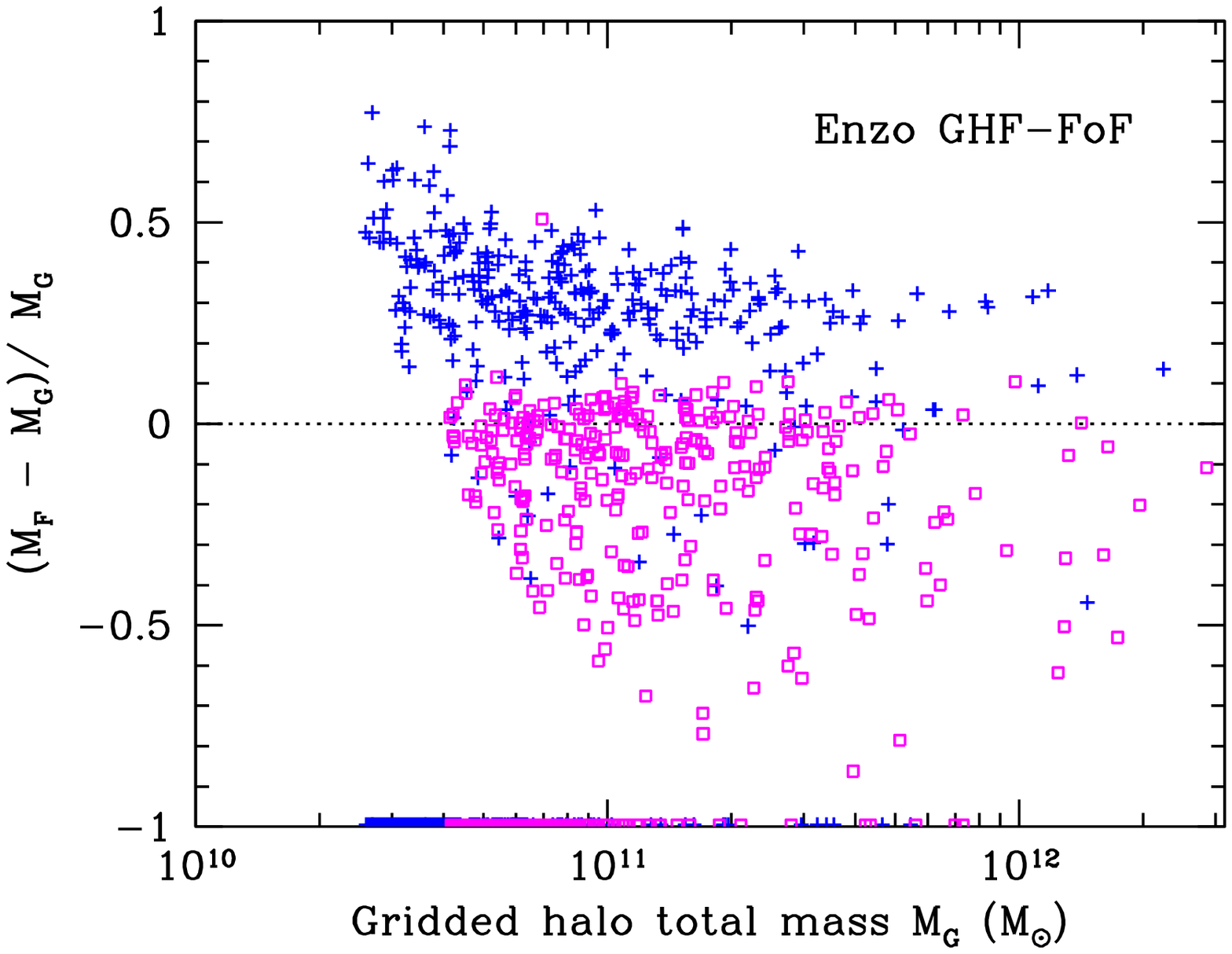}}
\vspace{-1.3cm}
\caption{Fractional halo mass differences between $N$-body haloes
  identified using FoF ($M_F$) and haloes found using the grid based
  halo finding algorithm ($M_G$), for the \texttt{Enzo} simulation
  E30$\_$512 at $z=3$. The blue crosses show the results when a
  threshold density of $\Delta_{\rm th}=178$ is used to identify the
  haloes in the grid halo finder, and the magenta open squares show
  results for $\Delta_{\rm th}=80$. The masses identified by the two
  approaches come into good agreement when the haloes on the gridded
  density field are grown to the lower overdensity threshold. The
  points along the bottom axis indicate unmatched haloes.
}
\label{fig:gMvsFM}
\end{figure}

Finally, Fig.~\ref{fig:gMvsFM} compares the masses of the haloes
identified using the grid based halo finder for two different density
thresholds with the masses of the matching haloes identified with FoF.
This demonstrates that the increase in the counts of the FoF haloes in
a given mass bin in Fig.~\ref{fig:griddedHMF} compared with the GHF
haloes arises primarily from the greater extents of the FoF haloes.
These enclose more mass and so shift the haloes to a higher mass bin
than the corresponding GHF haloes found using $\Delta_{\rm
  th}=178$. As found when comparing FoF and HOP halo counts, the
scatter shows that the rankings of the haloes by mass is not
preserved.

\end{appendix}

\label{lastpage}

\end{document}